\documentclass[12pt]{article}
\usepackage{a4wide}

\usepackage{amssymb,amsmath,amsthm,amscd}
\usepackage[mathcal]{euscript}
\usepackage{graphicx}

\newcommand{\vd}[2]{\frac{\delta #1}{\delta #2}}   
\newcommand{\Ef}{E^\varphi}
\newcommand{\Kf}{K_\varphi}

\newcommand{\lP}{\ell_{\rm P}}
\newcommand{\md}{{\mathrm d}}
\newcommand{\vt}{\vartheta}
\newcommand{\vp}{\varphi}

\begin{document}
{\renewcommand{\thefootnote}{\fnsymbol{footnote}}
\hfill  IGC--09/6--4\\
\medskip
\begin{center}
{\Large\bf Non-marginal LTB-like models with\\ inverse triad corrections from loop quantum gravity}\vspace{1.5em}
Martin Bojowald\footnote{e-mail address: {\tt bojowald@gravity.psu.edu}},
Juan D.~Reyes\footnote{e-mail address: {\tt jdr234@psu.edu}}
\\
\vspace{0.5em}
Institute for Gravitation and the Cosmos,
The Pennsylvania State
University,\\
104 Davey Lab, University Park, PA 16802, USA\\
\vspace{0.5em}
and Rakesh Tibrewala\footnote{e-mail address: {\tt rtibs@mailhost.tifr.res.in}}\\
\vspace{0.5em}
Tata Institute of Fundamental Research,\\
Homi Bhabha Road, Mumbai 400 005, India
\vspace{1.5em}
\end{center}
}

\begin{abstract}
 Marginal LTB models with corrections from loop quantum gravity have
 recently been studied with an emphasis on potential singularity
 resolution. This paper corroborates and extends the analysis in two
 regards: (i) the whole class of LTB models, including non-marginal
 ones, is considered, and (ii) an alternative procedure to derive
 anomaly-free models is presented which first implements
 anomaly-freedom in spherical symmetry and then the LTB conditions
 rather than the other way around. While the two methods give slightly
 different equations of motion, not altogether surprisingly given the
 ubiquitous sprawl of quantization ambiguities, final conclusions
 remain unchanged: Compared to quantizations of homogeneous models,
 bounces seem to appear less easily in inhomogeneous situations, and
 even the existence of homogeneous solutions as special cases in
 inhomogeneous models may be precluded by quantum effects. However,
 compared to marginal models, bouncing solutions seem more likely with
 non-marginal models.
\end{abstract}

\section{Introduction}

Quantum gravity changes the structure and dynamics of space-time on
small distance scales, which should have implications for the final
stages of matter collapse. An interesting class of models to shed
light on this issue is given by Lema\^itre--Tolman--Bondi (LTB)
space-times, which are inhomogeneous but do not show too much
complexity. Classically, these models describe collapsing dust balls,
containing Friedmann--Robertson--Walker solutions as
special cases. They thus provide an interesting extension of models
beyond homogeneity, an extension which is particularly important to
understand in the case of quantum gravity.

Loop quantum gravity implies characteristic correction terms in the
Hamiltonian constraint of gravity and matter. Once a combination of
corrections keeping the algebra of constraints anomaly-free has been
found, a canonical analysis of quantum gravitational collapse becomes
possible. Finding such equations is not easy, making the investigation
of implications from quantum gravity corrections highly
restricted. Quantum geometry corrections result from an underlying
spatial discreteness, which may cast doubt on whether they can leave
the theory covariant. There are arguments at the level of the full
theory \cite{QSDI} stating that quantum operators may be anomaly-free,
but it remains unknown how to descend from this statement to
anomaly-free effective space-time geometries. A phenomenological
approach has thus been followed to investigate possible geometrical
and physical effects of diverse corrections. Here, one inserts
expected corrections in the classical constraints, suitably
parameterized to reflect quantization ambiguities, and evaluates
conditions under which the corrected constraints remain first
class. As several articles have by now shown, it is indeed possible to
have anomaly-freedom even in the presence of quantum corrections
resulting from spatial discreteness
\cite{ConstraintAlgebra,LTB,SphSymmPSM}.
 
Marginal models, which are a subclass of general LTB models, have been
analyzed in this spirit in Ref.~\cite{LTB}. This has resulted in
consistent deformations which implement some types of quantum
corrections without spoiling general covariance, and made possible an
initial analysis of implications regarding effective pictures of
collapse singularities. (At the fundamental level of dynamical
difference equations in a loop quantization, spherically symmetric
models are singularity-free \cite{SphSymmSing} as are homogeneous
models \cite{Sing,IsoCosmo,HomCosmo}.) It turned out that there is no
clear generic avoidance of either space-like or null singularities by
an obvious mechanism, in contrast to several homogeneous models of
loop quantum cosmology \cite{LivRev} where phenomenological mechanisms
such as bounces could be found easily. While this outcome is not
entirely unexpected given the types of corrections analyzed in the
marginal case, it does show that further analysis is required.
Marginal models, after all, provide spatially flat
Friedmann--Robertson--Walker models in the homogeneous limiting case
which give rise to phenomenological singularity avoidance in their
loop quantization (including a positive matter potential) only with
holonomy corrections \cite{GenericBounce, APS}, which were not fully
included in \cite{LTB} due to technical complications. It is thus
natural to extend the constructions to non-marginal models which would
provide a homogeneous model of positive (as well as negative) spatial
curvature as limit. In that case, loop quantum cosmology can give rise
to phenomenological singularity resolution even in the presence of
inverse triad corrections alone \cite{BounceClosed}, which in
inhomogeneous situations are easier to control than holonomy
corrections. If the behavior seen in homogeneous models should be
generic and apply also to inhomogeneous situations, loop quantized
non-marginal LTB models must give rise to singularity resolution more
easily than marginal ones.

To find anomaly-free versions of non-marginal LTB models including
inverse triad corrections from loop quantum gravity, we will follow
two derivations. First, we will extend the methods of \cite{LTB} where
constraints already incorporating the LTB reduction of metric
components are made anomaly-free by consistency conditions between
correction functions. Secondly, we will derive a general anomaly-free
system of spherically symmetric constraints, on which we then apply
the LTB reduction in a second step. As we will show, the two steps of
LTB reduction and deriving consistency conditions almost commute: in
the end, we obtain consistent equations of motion of similar
structure, although they do differ by some terms. This outcome
considerably supports the constructions of \cite{LTB}.

Using these consistent equations, gravitational collapse can be
analyzed. We are specifically interested here in the possibility of a
turn-around of the collapse, or a bounce, in the corrected equations,
which are suggested to exist by models where homogeneous interiors
have been matched, Oppenheimer--Snyder-style, to spherically symmetric
exteriors \cite{Collapse}. Also here, as in the marginal case but in
contrast to homogeneous models, we do not find a clear indication for
singularity resolution, although several extra terms do seem to make a
bounce more likely. As in the marginal case, this part of the result
is not conclusive since not all corrections have been included and no
complete analysis has been performed. Our results thus do not mean
that there is no bounce in these inhomogeneous models. But they do
show that an outright treatment of inhomogeneous models is different
from matching homogeneous results. In fact, we also confirm the
observation of \cite{LTB} that quantum corrections of the type studied
here prevent the existence of an exact homogeneous
limit. ``Effective'' homogeneous geometries thus have to be taken with
care, but consistent relationships with inhomogeneous ones do provide
insights in their structure.

\section{Classical equations}

Non-marginal LTB models \cite{Lemaitre,TolmanSol,Bondi} have a
space-time metric given by
\begin{equation} \label{metric}
\md s^{2}=-\md t^{2}+\frac{R'^{2}}{1+\kappa(x)}\md x^{2}+R^{2}\md\Omega^{2}
\end{equation}
with $\md\Omega^2=\md\vt^2+\sin^2\vt\md\vp^2$ and where $\kappa\not=0$ is a function of the radial coordinate $x$. (The
limiting case $\kappa=0$ is that of marginal models.) The function
$R(t,x)$ can depend on both time and the radial coordinate, but not on the
angular coordinates to leave the metric spherically symmetric. It is
easy to see that positively curved Friedmann--Robertson--Walker models
with scale factor $a(t)$ are obtained for $\kappa(x)=-x^2$ and
$R(t,x)=a(t)x$.

For an application of loop quantization we use densitized triads
instead of the spatial metric components, whose conjugate momenta are
given in terms of the Ashtekar-Barbero connection and extrinsic curvature components. 
Written as a densitized vector field taking values in $\mathfrak{su}(2)$ with basis $\tau_i$, the spherically symmetric densitized triad is
\[
E=E^x(x)\tau_3\sin\vt\frac{\partial}{\partial x}+
(E^1(x)\tau_1+E^2(x)\tau_2)\sin\vt\frac{\partial}{\partial\vt}
+(E^1(x)\tau_2-E^2(x)\tau_1)\frac{\partial}{\partial\vp} \,.
\]
Similarly the Ashtekar connection $A^{i}_{a}=\Gamma^{i}_{a}+\gamma
K^{i}_{a}$, where $\Gamma^{i}_{a}$ and $K^{i}_{a}$ are the components
of spin connection and extrinsic curvature respectively and $\gamma$
is the Barbero-Immirzi parameter \cite{AshVarReell,Immirzi}, reads
\begin{eqnarray}
A&=&A_x(x)\tau_3\md x+(A_1(x)\tau_1+A_2(x)\tau_2)\md\vt \notag\\
&&+(A_1(x)\tau_2-A_2(x)\tau_1)\sin\vt\md\vp+ \tau_3\cos\vt\md\vp \notag
\end{eqnarray}
Introducing the U(1)-gauge invariant quantities
$(E^{\vp})^2=(E^1)^2+(E^2)^2$ and $A_{\vp}^2=A_1^2+A_2^2$, (see
\cite{SymmRed,SphSymm,SphSymmHam} for details) we have the
symplectic structure
\[
\{A_x(x),E^x(y)\}=\{\gamma K_\varphi(x),2E^\varphi(y)\}=
\{\eta(x),P^\eta(y)\}=2G\gamma \delta(x,y)
\]
or more explicitly the Poisson bracket of functions $f$ and $g$ is
\begin{align}
\{f,g\}=2G\int \md x\bigg(&\gamma\vd{f}{A_x}\vd{g}{E^x}+\frac{1}{2}\vd{f}{\Kf}\vd{g}{\Ef}+\gamma\vd{f}{\eta}\vd{g}{P^\eta}  \notag\\
             &-\gamma\vd{f}{E^x}\vd{g}{A_x}-\frac{1}{2}\vd{f}{\Ef}\vd{g}{\Kf}-\gamma\vd{f}{P^\eta}\vd{g}{\eta}\bigg)\,. \notag
\end{align}
Compared to metric variables, we have an extra field $\eta(x)$ with
momentum
\[
 P^{\eta}(x)=2A_{\varphi}E^{\varphi}\sin\alpha=4{\rm
 tr}\left((E^1\tau_1+E^2\tau_2)(A_2\tau_1-A_1\tau_2)\right)
\]
(with $\alpha$ defined as the angle between the internal directions of
$A$- and $E$-components), which plays the role of a U(1)-gauge angle
in the spherically symmetric theory. (This gauge angle also determines
the $x$-component of the spin connection $\Gamma_x=-\eta'$, and thus
enters the Ashtekar connection by $A_x=-\eta'+\gamma K_x$ with an
extrinsic curvature component $K_x$.)

In this situation, we have three constraints: the
Gauss constraint
\begin{equation}
G_{\rm grav}[\lambda]=\frac{1}{2G\gamma}\int \md x\,
\lambda((E^x)'+P^\eta)  \label{Gauss}
\end{equation}
the vector constraint
\begin{align}
D_{\rm grav}[N^x]&=\frac{1}{2G}\int \md x\,N^x\left(2E^\varphi K_\varphi'-\frac{1}{\gamma}A_x(E^x)'+\frac{1}{\gamma}\eta'P^\eta\right) \notag\\
        &=\frac{1}{2G}\int \md x\,N^x\left(2E^\varphi K_\varphi'-K_x(E^x)'+\frac{1}{\gamma}\eta'((E^x)'+P^\eta)\right)      \label{diffeomorphism}
\end{align}
and the Hamiltonian constraint 
\begin{equation} \label{Hamclass}
H_{\rm grav}[N]=-\frac{1}{2G}\int \md x\,N|E^x|^{-\frac{1}{2}}(K_\varphi^2E^\varphi+2K_\varphi K_xE^x + (1-\Gamma_\varphi^2)E^\varphi+2\Gamma_\varphi'E^x)
\end{equation}
with $\Gamma_\varphi=-(E^x)'/2E^\varphi$ the gauge invariant angular
component of the spin connection. Solving the Gauss constraint removes
the pair $(\eta,P^{\eta})$ and reduces the vector constraint to the
diffeomorphism constraint. After this step we can work with the canonical
pairs
\[
\{K_x(x),E^x(y)\}=\{K_\varphi(x),2E^\varphi(y)\}=2G \delta(x,y)\,.
\]

The relation to the usual spherically symmetric geometrodynamical
variables
\begin{equation} \label{geometrodynamicalVars}
\{R(x),P_R(y)\}=\{L(x),P_L(y)\}=G\delta(x,y)
\end{equation}
as used for example in \cite{LTBADM2,LTBADMnonmarg} can be obtained directly by
comparing the spatial metric 
\begin{equation} \label{spatialmetric}
 \md q^2= L^2\md x^2+R^2\md\Omega^2= \frac{(E^{\varphi})^2}{|E^x|}
\md x^2+|E^x|\md\Omega^2 
\end{equation}
in each set of variables and making use
of the equations of motion: 
\begin{eqnarray}
L=E^\varphi |E^x|^{-\frac{1}{2}}\quad&,&\quad R=|E^x|^\frac{1}{2}\,,\nonumber \\
P_L=-K_\varphi|E^x|^\frac{1}{2}\quad&,&\quad P_R=-sK_x |E^x|^\frac{1}{2} - K_\varphi E^\varphi |E^x|^{-\frac{1}{2}} \label{canonicalTransform}
\end{eqnarray}
where $s={\rm sgn}(E^x)$. (The sign factor corresponds to the two
possible orientations of a triad; we will mostly use $s=+1$ below.)
 
Specializing the general spherically symmetric metric
\[
\md s^2=-N(t,x)^2\md t^2+L^2(t,x)(\md x+N^x(t,x)\md t)^2 
+R^2(t,x)\md\Omega^2
\]
to the LTB form \eqref{metric} requires a vanishing shift function
$N^x=0$ and lapse $N=1$ for comoving coordinates of the dust, and
on using the first equation in \eqref{canonicalTransform} gives the
non-marginal LTB condition
\begin{equation} \label{ltb cond triad}
2\sqrt{1+\kappa(x)}E^{\varphi}=(E^x)'
\end{equation}
 in terms of triads. From this we can derive the spin connection
component
\begin{equation}\label{GammaNonMarg}
 \Gamma_{\varphi}=-\frac{(E^x)'}{2E^{\varphi}}=-\sqrt{1+\kappa(x)}
\end{equation}
and its derivative
$\Gamma'_{\varphi}=-\kappa'(x)/2\sqrt{1+\kappa(x)}$, which
 appear in the Hamiltonian constraint. Since the spin connection,
unlike in the marginal case, is not a constant $-1$, the Hamiltonian
constraint is different from the marginal case:
\begin{equation} \label{hamiltonian}
H_{\rm grav}^{\rm class}[N]=-\frac{1}{2G}\int
\md xN(x)|E^{x}|^{-1/2}\left(K_{\varphi}^{2}E^{\varphi}+2K_{\varphi}K_{x}E^{x}-\kappa(x)E^{\varphi}-\frac{\kappa'(x)E^{x}}{\sqrt{1+\kappa(x)}}\right)\,.
\end{equation}

If we solve the diffeomorphism constraint identically, which requires
$2E^\varphi K_\varphi'-K_x(E^x)'=0$, the LTB condition for triad variables
gives rise to a condition
\begin{equation} \label{ltb cond curvature}
K_{\varphi}'=\sqrt{1+\kappa(x)}K_{x}
\end{equation}
for the extrinsic curvature components. For a consistent LTB
formulation, the two LTB conditions must be preserved by evolution
generated by the constraint. This is indeed the case as can be seen
from deriving Poisson brackets between the Hamiltonian constraint and
each of the LTB conditions.  For the Poisson bracket of
the two LTB conditions, after smearing them with fields $\mu(x)$ and
$\nu(x)$, we get
\begin{eqnarray}
&& \left\{\int \md x\nu(x)\left(\sqrt{1+\kappa(x)}K_{x}-K_{\varphi}'\right),\int
\md y\mu(y)\left(2\sqrt{1+\kappa(x)}E^{\varphi}-(E^x)'\right)\right\} \nonumber \\
&& =2G\int \md z\sqrt{1+\kappa(z)}(\mu\nu)'\,.
\end{eqnarray}  
This in general is non-vanishing, unlike in the marginal case where
$\kappa$ is zero. (Although we will not follow
this route here, we note that this will have an impact on implementing
the LTB conditions at the state level, as done in \cite{LTB} for the
marginal case. Another complication for such a construction is the
explicit $\kappa$-dependence of the LTB conditions, which makes their
integrated version used as conditions on holonomies more complicated.)

Equations of motion in this canonical formulation are derived using
$\dot{E}^{x}=\{E^{x},H_{\rm grav}^{\rm class}\}$, with a similar
equation for $E^{\varphi}$. With these we can first eliminate $K_{x}$
and $K_{\varphi}$, and finally $E^{\varphi}$ using the non-marginal LTB
condition to obtain an equation entirely in terms of $E^{x}$. After
replacing $E^{x}$ by $R^{2}$ we obtain
\begin{equation}
H_{\rm grav}^{\rm class}=\frac{-2R\dot{R}\dot{R'}-\dot{R}^{2}R'+\kappa'R+\kappa R'}{2G\sqrt{1+\kappa(x)}} \,,
\end{equation}
which has to be equated to the matter part of the Hamiltonian for dust
given by $H_{\rm dust}=-\frac{1}{2G}F'/\sqrt{1+\kappa(x)}$. (A more
general canonical derivation of the gravity-dust system will be given in
Sec.~\ref{s:Dust}.)  Thus
\begin{equation} \label{first order eq classical}
2R\dot{R}\dot{R'}+\dot{R}^{2}R'-\kappa'R-\kappa R'=F'
\end{equation}
is the equation of motion, in agreement with the spatial derivative of
$R\dot{R}^{2}=\kappa(x)R+F(x)$, which is the equation obtained by
solving Einstein's equation for the non-marginal case.

\section{Inverse triad corrections from loop quantum gravity}

We will now repeat the canonical analysis using a Hamiltonian
constraint containing correction functions as they are suggested by
constraint operators in loop quantum gravity. Consistency will then
require conditions for the possible terms, which show
how quantum corrections can be realized in an anomaly-free way. We
discuss here only inverse triad corrections which are easier to
implement, and which already provide insights into one of the main
classes of quantum geometry corrections.

Inverse triad corrections arise from every Hamiltonian operator
quantized by loop techniques, where inverse components of the
densitized triad appear. Such corrections are directly related to
spatial discreteness of quantum geometry since densitized triads as
basic variables are quantized to flux operators with discrete spectra
containing zero \cite{LoopRep}. Since such operators do not have
densely defined inverses, no direct inverse operator is
available. Instead, well-defined quantizations exist based on
techniques introduced in \cite{QSDI,QSDV}, implying corrections to the
classical inverse.

\begin{figure}
\begin{center}
\includegraphics[width=16cm]{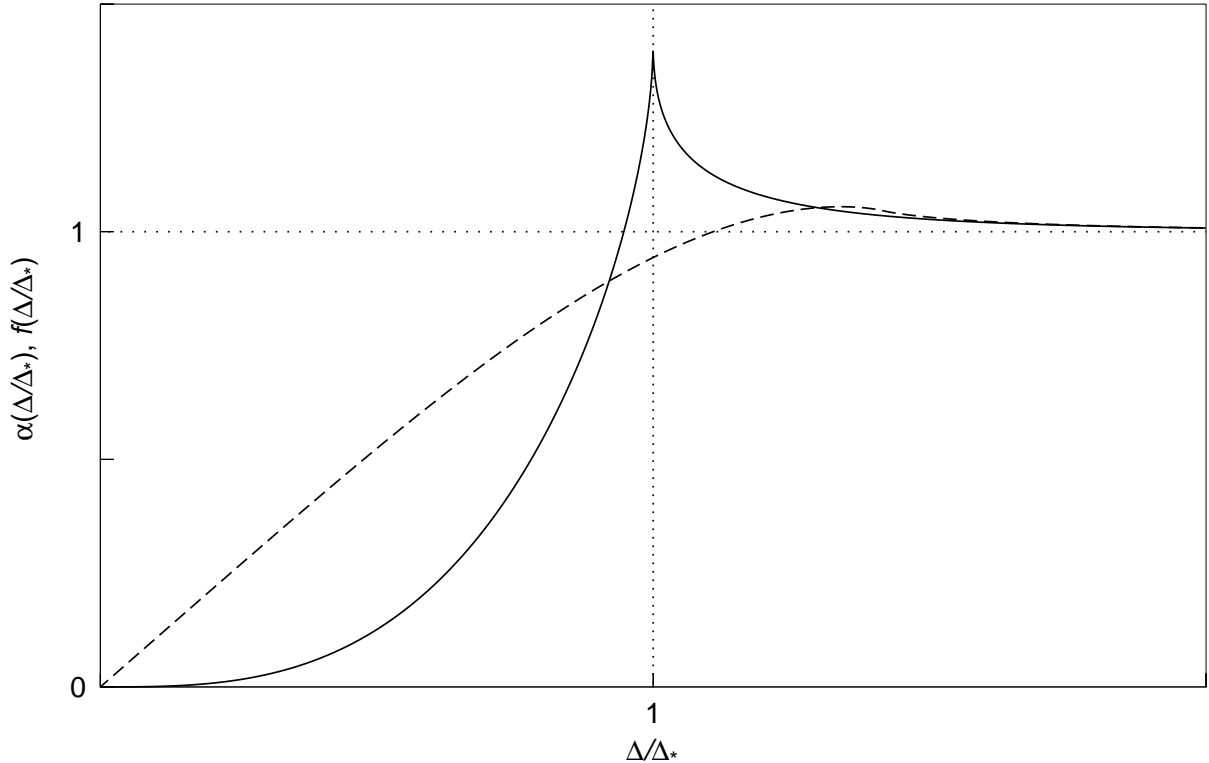}
\caption{\label{LTBalpha} The correction functions $\alpha(\Delta)$ (solid line) and $f(\Delta)$ (dashed line) where $\Delta$ is taken relative to $\Delta_*:=\sqrt{\gamma/2}\ell_{\rm P}$.}
\end{center}
\end{figure}

In several symmetric models, inverse triad operators and the
corrections they imply can be computed explicitly \cite{InvScale}. As
an example, spherically symmetric models used in \cite{LTB} give rise
to a correction function of the form
\begin{equation} \label{alpha}
\alpha(\Delta)=2\frac{\sqrt{|\Delta+\gamma \lP^{2}/2|}-
\sqrt{|\Delta-\gamma \lP^{2}/2|}}{\gamma \lP^{2}}\sqrt{|\Delta|}
\end{equation}
where $\Delta$ is the size of an elementary plaquette in a discrete
state underlying an LTB geometry. For corrections in inverse powers of
$E^x$, which is proportional to the area of a spherical orbit, the
relevant operators give rise to a dependence on plaquette sizes on
orbits. For a nearly spherical distribution of ${\cal N}(E^x)$ such
plaquettes making up the whole orbit, we thus have $\Delta=E^x/{\cal
N}$. Since we refer only to the orbit size, corrections thus naturally
depend on $E^x$ only but not on $E^{\varphi}$. (That this is required
will later be shown independently when we use anomaly freedom to rule
out that $\alpha$ could depend on $E^{\varphi}$.)  Such a correction
function then multiplies any classical appearance of $(E^x)^{-1}$ in a
Hamiltonian operator. In particular, classical divergences of inverse
factors of $E^x$ are cut off as one can see from the plot in
Fig.~\ref{LTBalpha}. Correspondingly, the dynamics given by such a
Hamiltonian will change from quantum corrections.  Classically, i.e.\
for $\lP\to0$, we have $\alpha(E^x)=1$, and this limit is approached
for large $E^x$. Also this behavior of the correction function is
illustrated in Fig.~\ref{LTBalpha}.

For a large number ${\cal N}$ of discrete blocks, the scale $\Delta$
is reduced compared to $E^x$ and corrections from $\alpha$ can be
significant even for large $E^x$. Since typically ${\cal N}$, in
relation to the underlying state, is not a constant but would depend
on the size $E^x$, different kinds of behaviors can arise. This
phenomenon of lattice refinement \cite{InhomLattice,SchwarzN} is
important to capture the full dynamics of quantum gravity and its
elementary degrees of freedom. It is also crucial for realizing the
correct scaling behavior of correction terms under changes of
coordinates \cite{Consistent}.  In this paper, we will mainly be
looking at general implications in local equations of motion, where
the value or behavior of ${\cal N}$ is not important. More detailed
investigations could at some point provide restrictions on the
possible form of ${\cal N}(E^x)$, and thus give insights in the
required behavior of discrete quantum gravity states.

\subsection{First version}

We turn to the case of inverse triad corrections called 'first
version' in \cite{LTB}, where only those terms in the Hamiltonian with
explicit $1/\sqrt{|E^{x}|}$ dependence are corrected by a factor
$\alpha(E^x)$. Starting as in the marginal case, we first assume that the
classical expression for the spin connection can be used, but show
that this is inconsistent. The Hamiltonian, now assuming $E^x>0$, is
\begin{equation} \label{hamiltonian first version}
H_{\rm grav}^{I}[N]=-\frac{1}{2G}\int \md x\,N\left(
\alpha(E^{x})\frac{K_{\varphi}^{2}E^{\varphi}}{\sqrt{E^{x}}}+
2K_{\varphi}K_{x}\sqrt{E^{x}}-\alpha(E^{x})\frac{\kappa(x)E^{\varphi}}{\sqrt{E^{x}}}-
\frac{\kappa'(x)\sqrt{E^{x}}}{\sqrt{1+\kappa(x)}}\right)\,.
\end{equation}

As in the marginal case, it turns out that the correction in the
Hamiltonian can lead to consistent LTB-type solutions only if we also
change the LTB conditions by a correction function $f(E^{x})$:
\begin{equation} \label{LTB1}
(E^x)'=2\sqrt{1+\kappa(x)}f(E^{x})E^{\varphi} \quad\mbox{ and }\quad
K_{\varphi}'=\sqrt{1+\kappa(x)}f(E^{x})K_{x}\,.
\end{equation} 
These relations still solve the classical diffeomorphism constraint identically, which does not receive corrections in loop quantum gravity.  The main consistency condition then is that Poisson brackets of the LTB conditions with the
Hamiltonian constraint vanish. 

Each of the Poisson brackets gives a differential equation for
$f(E^{x})$: For the LTB condition corresponding to the triad variables
we obtain
\begin{equation} \label{diff eq for f from triads}
2 E^x\frac{\md f}{\md E^{x}}=f(1-\alpha)\,.
\end{equation}
This is the same equation as found for the marginal case in
\cite{LTB}. For $\mathcal{N}=1$, for instance, the solution is  given by
\begin{equation}
 f(E^{x})=\frac{c_1\sqrt{E^{x}}e^{-\alpha/2}}{\left(\sqrt{E^{x}}+
 \sqrt{E^{x}-\gamma \lP^{2}/2}\right)^{1/2} \left(\sqrt{E^{x}}+
 \sqrt{E^{x}+\gamma \lP^{2}/2}\right)^{1/2}}
\end{equation}
for $E^x>\gamma\ell_{\rm P}^2/2$ and
\begin{equation}
 f(E^{x})= \frac{c_2\sqrt{E^x} \exp\left(-\frac{1}{2}\alpha+\frac{1}{2}\arctan
 \left(\sqrt{E^x/(\gamma\ell_{\rm P}^2/2-E^x)}\right)\right)}{
 \sqrt{\sqrt{E^x}+\sqrt{E^x+\gamma\ell_{\rm P}^2/2}}}
\end{equation}
for $E^x<\gamma\ell_{\rm P}^2/2$ (dashed curve in
Fig.~\ref{LTBalpha}). Here $c_1=2\sqrt{e}$ and $c_2=
2^{5/4}e^{1/2-\pi/4}\gamma^{-1/4}\ell_{\rm P}^{-1/2}$ are constants of
integration fixed, respectively, by the condition
$\lim_{E^{x}\rightarrow\infty}f(E^{x})\rightarrow1$ and by requiring
that $f(E^{x})$ be continuous at $E^x=\gamma\ell_{\rm P}^2/2$.

It is clear that we obtain the same equation because there is no
influence of $\kappa$ in the evaluation of the Poisson bracket,
$\kappa$ not affecting the terms containing $K_{x}$ and $K_{\varphi}$
in the Hamiltonian.  However, the differential equation for $f(E^{x})$
obtained by demanding that the corrected LTB condition for extrinsic
curvature components is also preserved in time gives the equation
\begin{equation} \label{diff eq for f from connection}
-2\sqrt{1+\kappa}K_{\varphi}K_{x}\sqrt{E^{x}}\frac{\md f}{\md E^{x}}+
\frac{\sqrt{1+\kappa}K_{\varphi}K_{x}f}{\sqrt{E^{x}}}-
\frac{\alpha\sqrt{1+\kappa}K_{\varphi}K_{x}f}{\sqrt{E^{x}}}-
\frac{\kappa'f}{2\sqrt{E^{x}}}+\frac{\kappa'\alpha}{2\sqrt{E^{x}}}=0
\end{equation}
which, due to the $\kappa'$-terms, is different from and in fact
inconsistent with equation (\ref{diff eq for f from triads})
obtained from the LTB condition for triads.

\subsection{Second version}

We now repeat the above procedure for the second version of the inverse triad
corrections for which the Hamiltonian is corrected by $\alpha(E^{x})$ in all terms and reads
\begin{equation} \label{hamiltonian second version}
H_{\rm grav}^{II}[N]=-\frac{1}{2G}\int \md x\,N
\frac{\alpha(E^{x})}{\sqrt{E^{x}}}\left(K_{\varphi}^{2}E^{\varphi}+2K_{\varphi}K_{x}E^{x} -\kappa(x)E^{\varphi}-\frac{\kappa'(x)E^{x}}{\sqrt{1+\kappa(x)}}\right)\,.
\end{equation}
Again, the LTB conditions in the form
\begin{equation} \label{LTB2}
(E^x)'=2\sqrt{1+\kappa(x)}g(E^{x})E^{\varphi} \quad\mbox{ and }\quad
K_{\varphi}'=\sqrt{1+\kappa(x)}g(E^{x})K_{x}
\end{equation} 
solve the diffeomorphism constraint identically.  However, for the
same reasons as noted above the conditions cannot be consistent: since
the terms containing $K_{x}$ and $K_{\varphi}$ in the Hamiltonian do
not involve $\kappa$, the Poisson bracket for the triads gives the
same result as in the marginal case where the differential equation
was
\begin{equation} \label{diff eq for g from triads}
\alpha\frac{\md g}{\md E^{x}}=g\frac{\md\alpha}{\md E^{x}}\,,
\end{equation}
with the solution $g(E^{x})=\alpha(E^{x})$. (This solution is unique
with the boundary condition imposing that $g=1$ for large arguments.)

For the Poisson bracket involving the condition on extrinsic
curvature, on the other hand, the terms in the Hamiltonian
involving $\kappa$ are important and we get a different result:
\begin{eqnarray} \label{diff eq for g from connection}
&&-2\sqrt{1+\kappa(x)}\alpha K_{\varphi}K_{x}\sqrt{E^{x}}
\frac{\md g}{\md E^{x}}+2\sqrt{1+\kappa(x)}K_{\varphi}K_{x}\sqrt{E^{x}}
\frac{\md\alpha}{\md E^{x}}g-\frac{\kappa'\alpha
 g}{2\sqrt{E^{x}}} \nonumber \\
&&-\kappa'\sqrt{E^{x}}\frac{\md\alpha}{\md E^{x}}g+
\frac{\kappa'\alpha}{2\sqrt{E^{x}}}=0
\end{eqnarray}
As in the previous case the presence of $\kappa'$ terms spoils the
consistency: Using $\alpha=g$ the first two terms cancel while the
rest would require $\md\alpha/\md E^x=(\alpha-1)/2E^x$ with a solution
$\alpha=1+c\sqrt{E^x}$ violating the classical limit at large
arguments.

\subsection{Inclusion of corrections in the spin connection}

A direct extension of the results from marginal to non-marginal models
is thus impossible. Here we have an example for the information gained
by a phenomenological treatment: LTB-type solutions require additional
corrections to compensate inconsistencies seen so far. Such
corrections may be more difficult to derive from a full Hamiltonian,
but they follow directly from a phenomenological treatment. A
successful consistent implementation thus provides feedback on the
full theory: additional corrections required for consistency must
eventually follow from the full theory just like the primary
correction $\alpha$ followed from inverse triad operators.

In particular, to resolve inconsistencies, we thus have to include
further corrections in terms not affected yet, the chief candidate
being the spin connection terms in the Hamiltonian constraint. They
vanish in the marginal case, such that results from there do not
provide much directions for more general models. Moreover, such terms
are in fact more difficult to derive from a full Hamiltonian so that
not much is known about their form. We will now look for corrections
in the spin connection terms which are such that they combine with
those already used to provide a consistent formulation.

\subsubsection{Implementation}

Classically we started with the expression
$\Gamma_{\varphi}=-(E^x)'/2E^{\varphi}=-\sqrt{1+\kappa(x)}$. However,
with quantum corrections to the Hamiltonian the LTB conditions are
also corrected. For example for the second version the modified LTB
condition $(E^x)'=2\sqrt{1+\kappa}g(E^{x})E^{\varphi}$, implies that
now the spin connection is
\begin{equation} \label{corrected gamma function}
\Gamma_{\varphi}=-\frac{(E^x)'}{2E^{\varphi}}=-\sqrt{1+\kappa(x)}g(E^{x})\,.
\end{equation}
We include the additional factor of $g(E^x)$ in the Hamiltonian by
replacing any occurrence of the classical spin connection
$\Gamma_{\varphi}^{\rm class}$ from (\ref{GammaNonMarg}) with
$g(E^{x})\Gamma^{\rm class}_{\varphi}$.

Furthermore, the derivative of the spin connection then is
\begin{equation}
\Gamma_{\varphi}'=-\frac{\kappa'
  g(E^{x})}{2\sqrt{1+\kappa}}-\sqrt{1+\kappa}(E^{x})'\frac{\md g}{\md E^{x}}\,.
\end{equation}
This introduces an explicit $(E^x)'$ in the LTB-reduced Hamiltonian,
which would imply $\{H[N],H[M]\}\neq 0$ even though the diffeomorphism
constraint has been solved identically. The system would thus be
anomalous. We are finally led to incorporate another correction
function multiplying $\Gamma'_{\varphi}$ in a function $h(E^{x})$ so
that in the Hamiltonian $\Gamma_{\varphi}'$ is to be replaced with
$h(E^{x})(\Gamma_{\varphi}^{\rm class})'$. The form of $h(E^{x})$ will be determined
by the requirement of consistency.  With all the possible corrections,
the new Hamiltonian in the second version is
\begin{eqnarray} \label{spin corrected hamiltonian ver two}
H_{\rm grav}^{II}[N]&=&-\frac{1}{2G}\int \md x\,N
\Bigl(\frac{\alpha K_{\varphi}^{2}E^{\varphi}}{\sqrt{E^{x}}}+
2\alpha K_{\varphi}K_{x}\sqrt{E^{x}}+
\frac{\alpha E^{\varphi}}{\sqrt{E^{x}}}
\nonumber \\
& &-\frac{\alpha g^{2}E^{\varphi}}{\sqrt{E^{x}}}
-\frac{\kappa\alpha g^{2}E^{\varphi}}{\sqrt{E^{x}}}-
\frac{\kappa'\alpha h\sqrt{E^{x}}}{\sqrt{1+\kappa}}\Bigr)\,.
\end{eqnarray}

We now demand that this Hamiltonian Poisson commutes with the LTB
conditions in \eqref{LTB2}.  Here we note that the terms containing
the spin connection (and its derivative) in the Hamiltonian do not
contain $K_{x}$ or $K_{\varphi}$ and therefore in evaluations of the
Poisson bracket with the first LTB condition there
will be no changes. This leads to the same differential equation (\ref{diff eq
for g from triads}) for $g(E^{x})$ as obtained earlier, implying
$g(E^{x})=\alpha(E^{x})$.  Evaluating the Poisson bracket of the
Hamiltonian with the second condition and using the solution for
$g(E^{x})$ gives a differential equation for $h(E^{x})$:
\begin{equation} \label{eq for h}
\frac{\md(\alpha h\sqrt{E^{x}})}{\md E^{x}}=\frac{\alpha^{2}}{2\sqrt{E^{x}}}\,.
\end{equation}
Here, we can thus have a consistent formulation for the non-marginal
case for a suitable $h$ by correcting the spin connection terms.

We proceed in a similar manner for the first version of the inverse
triad corrections. From the LTB condition for triads in \eqref{LTB1} we
find that the spin connection would be replaced with
$f(E^{x})\Gamma_{\varphi}^{\rm class}$, and the derivative of the spin connection
receives a correction function $l(E^{x})$ in the form
$l(E^{x})(\Gamma_{\varphi}^{\rm class})'$. With these changes the Hamiltonian is
\begin{eqnarray} \label{spin corrected hamiltonian ver one}
H_{\rm grav}^{I}[N]&=&-\frac{1}{2G}\int \md x\,N
\Bigl(\frac{\alpha K_{\varphi}^{2}E^{\varphi}}{\sqrt{E^{x}}}+
2 K_{\varphi}K_{x}\sqrt{E^{x}}+
\frac{\alpha E^{\varphi}}{\sqrt{E^{x}}}
\nonumber \\
& &-\frac{\alpha f^{2}E^{\varphi}}{\sqrt{E^{x}}}-
\frac{\kappa\alpha f^{2}E^{\varphi}}{\sqrt{E^{x}}}-
\frac{\kappa'l\sqrt{E^{x}}}{\sqrt{1+\kappa}}\Bigr)\,.
\end{eqnarray}
Evaluating the Poisson bracket of this with the LTB conditions and equating
them to zero implies that the equation for $f(E^{x})$ is unchanged compared
to (\ref{diff eq for f from triads}). The other Poisson bracket then gives a
differential equation for $l(E^{x})$:
\begin{equation} \label{eq for l}
\frac{\md(l\sqrt{E^{x}})}{\md E^{x}}=\frac{\alpha f}{2\sqrt{E^{x}}}\,.
\end{equation}

Differential equations for the correction functions $h$ and $l$ are
difficult to solve in general for given $\alpha$ and $f$.  For a near
center analysis done later we will need the lowest term in $h$ and $l$ in a
power series expansion in $x$. Integrating \eqref{eq for h} and \eqref{eq for l} keeping only the lowest order term in $\alpha$ and $f$ we find the solution
\begin{equation} \label{l near center}
h(x\approx0)=\left(\frac{2}{\gamma \lP^{2}}\right)^{\frac{3}{2}}\frac{R^{3}}{7} \quad , \quad l(x\approx0)=\frac{8e^{\frac{1}{2}-\frac{\pi}{4}}}{5(\gamma \lP^{2})^{2}}R^{4}
\end{equation}
valid near the center.

\subsubsection{Ambiguities}

The $E^x$-dependence of $\alpha$ follows from the consideration of
inverse triad operators in the full quantum theory. Although it is not
determined uniquely in this way (see \cite{Ambig,Robust} for
a discussion), the general shape of this correction function is known
well. No such arguments exist for some other correction functions such
as $h$, whose form is thus less clear. More ambiguities are thus
expected to arise for it.

A basic condition on multiplicative corrections is that they be scalar
to preserve the transformation properties of corrected expressions
under changing coordinates. Among the basic triad variables, $E^x$ is
the only one free of a density weight and thus can appear in
correction functions in an unrestricted way. The other component
$E^{\varphi}$, on the other hand, is a density of weight one and would
have to appear in combination with other densities to result in a
scalar. If only triad components are considered for the dependence,
the only other density would be $(E^x)'$. Scalars made from these
densities, such as $E^{\varphi}/(E^x)'$ are however unsuitable for
corrections since they are not always finite. 

In the present situation, we use the function $\kappa$ for a
non-marginal LTB model, which means that we have another density,
$\kappa'$, at our disposal. Scalars of the form $(E^x)'/\kappa'$ or
$E^{\varphi}/\kappa'$ are well-defined for most functions $\kappa$ of
interest, and can thus arise in corrections. This enlargement of the
space of acceptable variables means that additional ambiguities can
arise. In the next section we will see how several of these
ambiguities can be fixed by an analysis of the constraint algebra. The
equations of motion will remain structurally similar, so that we
proceed for now with an analysis of the equations resulting from the
treatment done so far.

\subsection{Equations of motion}

Given the consistency conditions between correction functions we can
derive consistent equations of motion even without having explicit
solutions for the differential equations \eqref{eq for h} and
\eqref{eq for l}.  Once consistent constraints are available, the
derivation follows the classical lines which we briefly illustrate
first: The first order equation (in time) has already been worked out
in \eqref{first order eq classical}, so that we can go on to the evolution
equation. From $\dot{E}^{x}=\{E^{x},H_{\rm grav}^{\rm class}\}$ we have
\begin{equation} 
K_{\varphi}=\frac{\dot{E}^{x}}{2\sqrt{E^{x}}}\,.
\end{equation}
Similarly, using $\dot{K}_{\varphi}=\{K_{\varphi},H_{\rm grav}^{\rm
class}\}$ we obtain
\begin{equation}
\dot{K}_{\varphi}=\frac{1}{2}\left(\frac{\kappa}{\sqrt{E^{x}}}-\frac{K_{\varphi}^{2}}{\sqrt{E^{x}}}\right)\,.
\end{equation}
Eliminating $K_{\varphi}$ from the above two equations we obtain
\begin{equation}
\ddot{E}^{x}=\kappa+\frac{(\dot{E}^{x})^{2}}{4E^{x}}
\end{equation}
which, using $E^{x}=R^{2}$, can be written as
\begin{equation} \label{evolution eq classical}
2R\ddot{R}+\dot{R}^{2}=\kappa\,.
\end{equation}
Eqs.~(\ref{first order eq classical}) and (\ref{evolution eq
classical}) are automatically consistent, which can be seen explicitly
by subtracting a time derivative of (\ref{first order eq classical})
from a space derivative of (\ref{evolution eq classical}).

The same procedure is then applied to constrained systems including
consistent correction terms. To get the first order equation in
version two we use \eqref{spin corrected hamiltonian ver two} in the
equations of motion $\dot{E}^{a}=\{E^{a},H^{II}_{\rm grav}\}$ to
solve for the extrinsic curvature components
\begin{equation} \label{kphi dot}
 K_{\varphi}=\frac{\dot{E}^{x}}{2\alpha\sqrt{E^{x}}} \quad\mbox{ and }
\quad
K_{x}=\frac{\dot{E}^{\varphi}}{\alpha\sqrt{E^{x}}}-\frac{K_{\varphi}E^{\varphi}}{E^{x}}\,.
\end{equation}
Using these along with the LTB condition
$E^{\varphi}=(E^{x})'/2\sqrt{1+\kappa}g$ in \eqref{spin corrected
hamiltonian ver two} we rewrite the Hamiltonian in terms of $E^x$
only:
\begin{eqnarray} 
H^{II}_{\rm grav}&=&-\frac{1}{2G}\Bigl(-\frac{(\dot{E}^{x})^2(E^{x})'}{8\sqrt{1+\kappa}\alpha^{2}(E^{x})^{3/2}}+\frac{\dot{E}^{x}(\dot{E}^{x})'}{2\sqrt{1+\kappa}\alpha^{2}\sqrt{E^{x}}}-
\frac{(\dot{E}^{x})^2(E^{x})'}{2\sqrt{1+\kappa}\alpha^{3}\sqrt{E^{x}}}\frac{\md\alpha}{\md E^{x}} \nonumber \\
&&+
(1-\alpha^{2}-\kappa\alpha^{2})\frac{(E^{x})'}{2\sqrt{1+\kappa}\sqrt{E^{x}}}-
\frac{\kappa'\alpha h\sqrt{E^{x}}}{\sqrt{1+\kappa}}\Bigr) \,,
\end{eqnarray} 
where we have already used the condition $g=\alpha$. When equated to the dust Hamiltonian after using $E^{x}=R^{2}$
a first order equation in time ensues:
\begin{equation} \label{first order eq ver two}
\frac{\dot{R}^{2}R'}{\alpha^{2}}+\frac{2R\dot{R}\dot{R}'}{\alpha^{2}}-\frac{2R\dot{R}^{2}R'}{\alpha^{3}}\frac{\md\alpha}{\md R}+(1-\alpha^{2}-\kappa\alpha^{2})R'-\kappa'\alpha hR=F'\,.
\end{equation}

To obtain the evolution equation we use
\begin{equation}
\dot{K}_{\varphi}=\{K_{\varphi},H^{II}_{\rm grav}\}=
-\frac{1}{2}\frac{\alpha-\alpha g^{2}-\kappa\alpha g^{2}+
\alpha K_{\varphi}^{2}}{\sqrt{E^{x}}}
\end{equation}
together with $K_{\varphi}=\dot{E}^{x}/2\alpha\sqrt{E^{x}}$ from
\eqref{kphi dot}, such that
\begin{equation}
\ddot{E}^{x}-\frac{(\dot{E}^{x})^{2}}{4E^{x}}-\frac{(\dot{E}^{x})^{2}}{\alpha}\frac{\md\alpha}{\md E^{x}}=-\alpha^{2}(1-g^{2}-\kappa g^{2})\,.
\end{equation}
With $g=\alpha$ and $E^{x}=R^{2}$ this becomes
\begin{equation} \label{evolution eq ver two}
2R\ddot{R}+\left(1-2\frac{\md\log\alpha}{\md\log R}\right)
\dot{R}^{2}=-\alpha^{2}(1-\alpha^{2}-\kappa\alpha^{2})\,.
\end{equation}
It is easy to see that this equation has the correct classical limit
and (using \eqref{eq for h}) is consistent with the first order equation.

Proceeding in a similar manner for the first version of the inverse
triad correction we find that the first order equation is
\begin{equation} \label{first order eq ver one}
\alpha\dot{R}^{2}R'+2\dot{R}\dot{R}'R+\alpha(1-f^{2}-\kappa f^{2})R'-\kappa'flR=fF'
\end{equation}  
and the evolution equation
\begin{equation} \label{evolution eq ver one}
2R\ddot{R}+\alpha\dot{R}^{2}=-\alpha(1-f^{2}-\kappa f^{2})\,.
\end{equation}

Using \eqref{eq for l} along with \eqref{diff eq for g from triads} one can verify explicitly that the first order and the second order equations are consistent with each other.

\subsection{Effective density}
\label{s:effectiveDensity}

To interpret effects from correction terms it is often
useful to formulate them in terms of effective densities rather than
new terms in equations of motion. As in the marginal case, we use the
Misner-Sharp mass  defined by
\begin{equation}
m=\frac{R}{2}(1-\nabla_{A}R\nabla^{A}R)
\end{equation}
where $A=(1,2)$ corresponds to the $t-r$ manifold.
Writing the metric for spherical dust collapse as 
\begin{equation}
\md s^{2}=-\md t^{2}+L^{2}(t,x)\md x^{2}+R^{2}(t,x)\md\Omega^{2}
\end{equation}
the equation for the Misner-Sharp mass becomes
\begin{equation} \label{misner sharp mass}
m=\frac{R}{2}\left(1+\dot{R}^{2}-\frac{R'^{2}}{L^{2}}\right)\,.
\end{equation}
Classically $R'^{2}/L^{2}=1+\kappa$ and therefore the Misner-Sharp
mass is $m=(R\dot{R}^{2}-\kappa R)/2$. With an effective density
defined in terms of the Misner-Sharp mass by
\begin{equation} \label{effective density}
\epsilon_{\rm eff}=\frac{m'}{4\pi GR^{2}R'}
\end{equation}
we find that for the classical collapse it is
\begin{equation} \label{classical eff density}
\epsilon_{\rm eff}^{\rm class}=\frac{F'}{8\pi GR^{2}R'}\,.
\end{equation}
Here we have made use of the equation of motion $\dot{R}^{2}R=\kappa
R+F$ which is obtained from the Hamiltonian constraint. This is in
agreement with the 00 component of Einstein's equation, $G_{00}=8\pi
G\epsilon(t,x)$ where $\epsilon(t,x)$ is the dust density, implying
$\epsilon=F'/8\pi GR^{2}R'$. Thus, classically the effective density,
defined in terms of the Misner-Sharp mass, is the same as the dust
density. Moreover, as expected, the expressions for the two are
unchanged compared to those for the marginal case.

We now proceed in the same way to find the effective density for the
first version of the inverse triad correction. With the new LTB
condition $E^{\varphi}=(E^x)'/2\sqrt{1+\kappa}f(E^{x})$, the metric
coefficient $L\equiv E^{\varphi}/\sqrt{E^{x}}$ implies
$L=R'/\sqrt{1+\kappa}f(R)$ using $E^{x}=R^{2}$. Therefore the
Misner-Sharp mass as defined in \eqref{misner sharp mass} is now
\begin{equation} \label{ms mass first version}
m^{I}=\frac{R}{2}(1+\dot{R}^{2}-(1+\kappa)f^{2})\,.
\end{equation}
The corresponding effective density as implied by \eqref{effective
density} is
\begin{equation} \label{eff density first version}
\epsilon_{\rm eff}^{I}=\frac{1}{8\pi GR^{2}}\left(\frac{fF'}{R'}+
(\alpha-1)(3f^{2}+3\kappa f^{2}-\dot{R}^{2}-1)+
\frac{\kappa'flR}{R'}-\frac{\kappa'f^{2}R}{R'}\right)
\end{equation}
where we have made use of \eqref{diff eq for f from triads} after
substituting for $E^{x}$ in terms of $R$, and of \eqref{first order eq
ver one}. We note that this equation has the correct classical limit.

For the second version of the inverse triad correction we have
$E^{\varphi}=(E^x)'/2\sqrt{1+\kappa}g(E^{x})$, implying
$L=R'/\sqrt{1+\kappa}g(R)$ where as seen earlier
$g(R)=\alpha(R)$. With this the Misner-Sharp mass is
\begin{equation} \label{ms mass second version}
m^{II}=\frac{R}{2}(1+\dot{R}^{2}-(1+\kappa)g^{2})\,.
\end{equation}
Using the relation $\alpha'=R'\md\alpha/\md R$ (where the prime denotes
derivative with respect to $x$) along with \eqref{first order eq ver
two} we find that the effective density is
\begin{equation} \label{eff density second version}
\epsilon_{\rm eff}^{II}=\frac{1}{8\pi G}\left(\frac{\alpha^{2}F'}{R^{2}R'}+\frac{1-\alpha^{2}}{R^{2}}(1-\alpha^{2}-\kappa\alpha^{2})+\frac{2}{\alpha R}(\dot{R}^{2}-\alpha^{2}-\kappa\alpha^{2})\frac{\md\alpha}{\md R}+\frac{\kappa'\alpha^{3}h}{RR'}-\frac{\kappa'\alpha^{2}}{RR'}\right)
\end{equation}

As in the marginal case, these effective densities imply that the near
center expansion for the mass function $F$ can have different
behavior compared to the classical case, as discussed in
Sec.~\ref{s:NearCenter}. The matter contribution to the effective
density, as given by the first terms of \eqref{eff density first
version} and \eqref{eff density second version}, is the same as in the
marginal case.

\subsection{Quantum correction to the energy function $\kappa$?}
\label{s:effectiveKappa}

Physically one would expect that the energy function $\kappa$, which
is related to the velocity of the dust cloud, should also receive
corrections after including quantum effects. To derive those, we have
to find an independent definition of $\kappa$ referring only to the
constraints or evolution equations derived from them. One
possibility, in the classical case, is to use \eqref{evolution eq
classical} whose right hand side only contains the energy
function. Once brought into an analogous form, a corrected evolution
equation can directly be used to read off a corrected energy
function. Specifically for version one, where the evolution equation
is given by \eqref{evolution eq ver one}, the effective energy function
is
\begin{equation} \label{eff kappa ver one}
\kappa_{\rm eff}^{I}=\alpha f^{2}\kappa -\alpha(1-f^2)
\end{equation}
while for version two, where the evolution equation is given by
\eqref{evolution eq ver two}, the effective energy function becomes
\begin{equation} \label{eff kappa ver two}
\kappa_{\rm eff}^{II}=\alpha^{4}\kappa-\alpha^2(1-\alpha^2)\,.
\end{equation}
This correction in effect would imply that the near center expansion
for $\kappa$ can be different for the quantum corrected equations as
we will see when we come to the near center analysis below.

\section{Spherically symmetric constraints}

We have now several versions of consistent sets of equations of motion
for non-marginal LTB models including inverse triad corrections as
expected from loop quantum gravity. To make these equations
consistent, we had to introduce several correction functions in
different terms of the Hamiltonian constraint, which were then related
to each other by consistency conditions following from the requirement
that the LTB conditions be preserved. Since there is some freedom in
choosing the places and forms of corrections in the constraint as well
as the LTB conditions, one may question how reliable such an analysis
is regarding the structure of resulting equations of motion or
implications for gravitational collapse.  

Before analyzing corrected equations of motion further, we now present
an independent derivation which starts with a consistent set of
corrected spherically symmetric constraints, and then implements the
LTB reduction. As we will see, the structure of the resulting
equations is nearly unchanged, while much less assumptions about
different corrections are required. With these two procedures we thus
demonstrate the robustness of consistently including corrections at a
phenomenological level. Note that this would not have been possible
had we chosen to fix the gauge generated by the Hamiltonian constraint
in any way instead of dealing with the anomaly-issue head-on.

\subsection{Gravitational variables and constraints}

Quantum corrections due to inverse powers of the densitized triad are
introduced in the Hamiltonian constraint (\ref{Hamclass}) by functions which we
initially assume to be of the general form $\alpha(E^x, E^\varphi)$
and $\bar{\alpha}(E^x,E^\varphi)$ entering the Hamiltonian constraint as
\begin{align}
H_{\rm grav}^Q[N]=-\frac{1}{2G}\int \md x\,N \big( \alpha\,|E^x|^{-\frac{1}{2}}K_\varphi^2E^\varphi+&2s\bar{\alpha}\,K_\varphi K_x |E^x|^\frac{1}{2} + \alpha\,|E^x|^{-\frac{1}{2}}E^\varphi   \notag \\
-&\alpha_\Gamma\,|E^x|^{-\frac{1}{2}}\Gamma_\varphi^2E^\varphi+ 2s\bar{\alpha}_\Gamma\,\Gamma_\varphi'|E^x|^\frac{1}{2}\big)\,.   \label{Hamiltonian}
\end{align}
To account for possible corrections from the quantization of the spin
connection, as suggested by the previous analysis, we have also
introduced functions $\alpha_\Gamma(E^x,E^\varphi)$ and
$\bar{\alpha}_\Gamma(E^x,E^\varphi)$ in those terms. The only
restriction so far is that we have the same $\alpha$ in the first and
third term of the Hamiltonian constraint due to their common origin
from the inverse $|E^x|^{-1/2}$. The two main cases of interest here
are $\bar{\alpha}=1$ or $\bar{\alpha}=\alpha$, corresponding to two
versions of inverse triad corrections.

We now proceed to make the corrected constraints anomaly-free before
implementing LTB conditions. (For a similar analysis for dilaton
gravity, see \cite{SphSymmPSM}.) To ensure anomaly-freedom, we must
determine conditions under which the system of constraints, including
its corrections in the Hamiltonian constraint, remains first
class. Computing the Poisson bracket $\{H_{\rm grav}^Q[M],H_{\rm
grav}^Q[N]\}$ gives
\begin{align} \label{HHD}
\{H_{\rm grav}^Q[M],H_{\rm grav}^Q[N]\}=&D_{\rm grav}[\bar{\alpha}\bar{\alpha}_\Gamma|E^x|(\Ef)^{-2}(MN'-NM')]\notag\\
 &-G_{\rm grav}[\bar{\alpha}\bar{\alpha}_\Gamma|E^x|(\Ef)^{-2}(NM'-MN')\eta']  \notag \\
&+\frac{1}{2G}\int \md x\,(MN'-NM')(\bar{\alpha}\alpha_\Gamma-\alpha\bar{\alpha}_\Gamma)\frac{s K_\varphi (E^x)'}{E^\varphi}   \notag \\
&+\frac{1}{2G}\int \md x\,(MN'-NM')(\bar{\alpha}' \bar{\alpha}_\Gamma -\bar{\alpha}\bar{\alpha}_\Gamma')\frac{2K_\varphi |E^x|}{E^\varphi} \,.
\end{align}

For a first class algebra the last two terms, which are not related to
constraints, must vanish, providing conditions on the correction
functions. The vanishing of the last term implies
$\bar{\alpha}_\Gamma\propto \bar{\alpha}$, upon which the third term
gives $\alpha_\Gamma\propto \alpha$. Therefore to recover the
classical limit we must have:
\begin{equation}
\alpha_\Gamma= \alpha \quad,\quad
\bar{\alpha}_\Gamma= \bar{\alpha} \,.
\end{equation}
Thus, anomaly freedom requires corrections to the spin connection
terms to be only due to the inverse power of the densitized triad
factors they contain. This may look contradictory to what we derived
earlier, where additional correction functions such as $h$ were
needed. However, the previous case (where LTB conditions were used
instead of the diffeomorphism constraint) implicitly makes $h$
dependent on $(E^x)'$ as well: Comparing the correction terms we have
\[
-\frac{\kappa'}{\sqrt{1+\kappa}}\alpha h = 2\bar{\alpha}_{\Gamma}\Gamma_{\varphi}' = -2\bar{\alpha}_{\Gamma}
\left(\frac{1}{2}\frac{\kappa'}{\sqrt{1+\kappa}} g[E^x]+
\sqrt{1+\kappa} \frac{\md g[E^x]}{\md E^x} (E^x)'\right) 
\]
and we can write, using $g=\alpha$:
\[
 h= \bar{\alpha}_{\Gamma}+2\frac{1+\kappa}{\kappa'}
 \frac{\md\log\alpha}{\md E^x} (E^x)'\,.
\]
Thus, to match the current equations the correction function $h$ used
earlier must depend on $(E^x)'$, which has a density weight. (Similar
considerations apply to the correction function $l$.)  As the
expression demonstrates, this is made possible since in our earlier
procedure we had the function $\kappa$ at our disposal in addition to
the triad components. Its derivative $\kappa'$ provides an extra
density, which can be combined with $(E^x)'$ to provide a scalar
correction function. In the current setting, by contrast, we have not
yet introduced any such function by LTB conditions, and so a possible
dependence on $(E^x)'$ is more restricted. The new procedure of this
section is clearly less ambiguous, while the final results will be
very close. This again demonstrates the robustness.

To continue with the analysis of anomaly-freedom, we compute Poisson brackets:
\begin{align}
\{H^Q_{\rm grav}[N],D_{\rm grav}[N^x]\}=-H^Q_{\rm grav}&[N^xN'] \notag\\
-\frac{1}{2G}\int \md x\,N (N^x)'E^\varphi \Bigl(& \frac{\partial\alpha}{\partial E^\varphi}|E^x|^{-\frac{1}{2}}K_\varphi^2E^\varphi+2s\frac{\partial\bar{\alpha}}{\partial E^\varphi}K_\varphi K_x |E^x|^\frac{1}{2} + \frac{\partial\alpha}{\partial E^\varphi}|E^x|^{-\frac{1}{2}}E^\varphi   \notag \\
&-\frac{\partial\alpha}{\partial E^\varphi}|E^x|^{-\frac{1}{2}}\Gamma_\varphi^2E^\varphi+ 2s\frac{\partial\bar{\alpha}}{\partial E^\varphi}\Gamma_\varphi'|E^x|^\frac{1}{2}\Bigr) \,.
\end{align}
In the case $\bar{\alpha}=\alpha$,
\[
\{H_{\rm grav}^Q[N],D_{\rm grav}[N^x]\}=-H_{\rm grav}^Q[N^xN'-(\partial\log\alpha/\partial\Ef)\Ef N (N^x)']\,.
\]
The corrected Hamiltonian $H^Q_{\rm grav}$ transforms as a scalar only
if $\alpha$ is independent of $E^\varphi$ since $E^\varphi$ is the
only basic quantity of density weight one. However, the vacuum algebra
is first class even if $\alpha$ depends on $E^\varphi$. In contrast,
when $\bar{\alpha}=1$ (or more generally $\bar{\alpha}\neq\alpha$),
$\alpha$ must be independent of $E^\varphi$. (The case
$\alpha=\bar{\alpha}$ in vacuum is special because any such correction
could be absorbed in the lapse function, making the algebra formally
first class.)

In summary, for corrections $\alpha$ (and $\bar{\alpha}$) independent
of $E^\varphi$ we have
\begin{equation}
\{H_{\rm grav}^Q[N],D_{\rm grav}[N^x]\}=-H_{\rm grav}^Q[N^xN'] \notag 
\end{equation}
and
\begin{eqnarray} \label{DeformedAlg}
\{H_{\rm grav}^Q[M],H_{\rm grav}^Q[N]\}&=&D_{\rm grav}[\bar{\alpha}^2|E^x|(\Ef)^{-2}(MN'-NM')]\\
&&-G_{\rm grav}[\bar{\alpha}^2|E^x|(\Ef)^{-2}(MN'-NM')\eta']   \nonumber
\end{eqnarray}
To proceed, we will include matter in the form of dust as it is
assumed in LTB models.

\subsection{Dust}
\label{s:Dust}

For a full consistency analysis based on the constraint algebra we
have to use a dynamical formulation of the dust matter source, rather
than a phenomenological implementation via the dust profile $F(x)$.
It is convenient to use a canonical formulation for dust with
stress-energy tensor $T_{\alpha\beta}=\epsilon\, U_\alpha U_\beta$ as
developed in \cite{BrownKuchar}. The dust four-velocity is given by
the Pfaff form $U_\alpha=-\tau,_\alpha+W_kZ^k,_\alpha$, where as
canonical coordinates the dust proper time $\tau$ and comoving dust
coordinates $Z^k$ with $k=1,2,3$ appear. Their respective conjugate
momenta will be called $P$ and $P_k$. Matter contributions to the
diffeomorphism and Hamiltonian constraint read
\begin{align}
D_{\rm dust}[N^a]=&\int \md^3x\,N^a\tilde{D}_a=\int d^3x\,N^a(P\tau,_a+P_kZ^k,_a) \notag \\
H_{\rm dust}[N]=&\int \md^3x\, N\sqrt{P^2+q^{ab}\tilde{D}_a\tilde{D}_b}  \label{fHdust}\,.
\end{align}

Imposing spherical symmetry and using adapted coordinates $\Phi:=Z^1$, $Z^2=\vartheta$, $Z^3=\varphi$ the constraints become
\begin{align}
D_{\rm dust}[N^x]=&4\pi\int \md x\, N^x\left(P_\tau\tau'+P_\Phi\Phi'\right) \notag \\
H_{\rm dust}[N]=&4\pi\int \md x\, N\sqrt{P_\tau^2+\frac{|E^x|}{(E^\varphi)^2}(P_\tau \tau'+P_\Phi\Phi')^2}
\end{align}
with the remaining canonical pairs
\[
\{\tau,P_\tau\}=\{\Phi,P_\Phi\}=\frac{1}{4\pi}
\]
whose momenta $P_\tau$ and $P_\Phi$ are defined by the relations
$P=P_\tau\sin\vartheta$ (in terms of the $P$ of the full 3-dimensional
theory) and $P_\Phi=-P_\tau W_1$.

For non-rotating dust, as must be the case with spherical symmetry, the
constraints $P_k=0$ can be imposed by requiring that the dust motion
be described with respect to the frame orthogonal foliation, so that the
state does not depend on the frame variables $Z^k$. As a result
$P_\Phi$ is usually taken to be zero. However, we will not choose to
do so until we try to solve the equations of motion.

From the form of the Hamiltonian (\ref{fHdust}) in the full theory and
$q^{ab}=(\det E^c_k)^{-1}E^a_iE^b_i$, we can expect quantum corrections
$\beta[E^x,E^\varphi]$ from a quantization of inverse triads inside the
square root:
\[
H^Q_{\rm dust}[N]=4\pi\int \md x\, N\sqrt{P_\tau^2+\beta\frac{|E^x|}{(E^\varphi)^2}(P_\tau \tau'+P_\Phi\Phi')^2}\,.
\]
Also here, the form of $\beta$ will be restricted by the requirement
of anomaly freedom.

Adding the individual contributions, the diffeomorphism and corrected
Hamiltonian constraint for the gravity-dust system are $D[N^x]=D_{\rm
grav}[N^x]+D_{\rm dust}[N^x]$ and $H^Q[N]=H^Q_{\rm grav}[N]+H^Q_{\rm
dust}[N]$.  Now, the Poisson bracket for the matter part of the
Hamiltonian with the diffeomorphism constraint is
\[
\{H^Q_{\rm dust}[N],D[N^x]\}=-H^Q_{\rm dust}[N^xN']+\int \md x\,NN^x\,'\frac{\partial\beta}{\partial E^\varphi}\frac{|E^x|}{2E^\varphi}\frac{\tilde{D}_x^2}{\sqrt{P_\tau^2+\beta|E^x|(E^\varphi)^{-2}\tilde{D}_x^2}}
\]
with $\tilde{D}_x:=P_\tau\tau'+P_\Phi\Phi'$. The closure of
$\{H^Q[N],D[N^x]\}$ consistently imposes the condition that $\alpha$
and $\beta$ be independent of $E^\varphi$, upon which
\[
\{H^Q[N],D[N^x]\}=-H^Q[N^xN']\,.
\]
Finally,
\begin{align}
\{H^Q[N],H^Q[M]\}=&\,\{H^Q_{\rm grav}[M],H^Q_{\rm grav}[N]\}+\{H^Q_{\rm dust}[M],H^Q_{\rm dust}[N]\} \notag \\
 =&\,D_{\rm grav}[\bar{\alpha}^2|E^x|(\Ef)^{-2}(MN'-NM')]-
G_{\rm grav}[\bar{\alpha}^2|E^x|(\Ef)^{-2}(MN'-NM')\eta']   \notag \\ 
 +&\,D_{\rm dust}[\beta|E^x|(E^\varphi)^{-2}(MN'-NM')] \notag 
\end{align}
gives the relation
\begin{equation}
\beta=\bar{\alpha}^2\,.
\end{equation}
Note that the presence of matter makes this consistent deformation of
the classical constraint algebra non-trivial: corrections can no
longer be absorbed in the lapse function. We also point out that a
deformation of the constraint algebra is required to implement the
corrections consistently. This seems to be an interesting difference
to a reduced phase space quantization which is possible in this class
of models based on a deparameterization \cite{BKDustLTB}.

\subsection{LTB-like solutions}

Using the transformation equations (\ref{canonicalTransform}), the
quantum corrected Hamiltonian in ADM variables (\ref{geometrodynamicalVars})
reads 
\begin{align}
H^Q[N]=\frac{1}{G}\int \md x\,N\bigg[-\bar{\alpha}\frac{P_LP_R}{R}+&(2\bar{\alpha}-\alpha)\frac{LP_L^2}{2R^2}-\alpha\frac{L}{2}-(2\bar{\alpha}-\alpha)\frac{R'\,^2}{2L}+\bar{\alpha}\left(\frac{RR'}{L}\right)' \notag \\
&+4\pi G P_\tau\left(1+\beta\frac{\tau'\,^2}{L^2}\right)^\frac{1}{2}\bigg] \notag
\end{align}
and the diffeomorphism constraint is
\[
D[N^x]=\frac{1}{G}\int \md x\,N^x(R'P_R-LP_L'+4\pi G P_\tau\tau')
\]
where we have already used $P_\Phi=0$. Paralleling our treatment of
LTB-reduced constraints our further analysis will be split into two
different cases of correction functions.  We choose to work here in
ADM variables instead of triad variables, but of course identical
results follow using the latter.

\subsubsection{Case $\bar{\alpha}=\alpha$}

The equations of motion $\dot{R}=\{R,H^Q[N]+D[N^x]\}$, $\dot{L}=\{L,H^Q[N]+D[N^x]\}$, $\dot{P_R}=\{P_R,H^Q[N]+D[N^x]\}$, $\dot{P_L}=\{P_L,H^Q[N]+D[N^x]\}$, $\dot{\tau}=\{\tau,H^Q[N]+D[N^x]\}$ and $\dot{P_\tau}=\{P_\tau,H^Q[N]+D[N^x]\}$ are respectively,
\begin{align}
P_L&=\frac{R}{\alpha N}\left(-\dot{R}+N^xR'\right) \label{eqPL}\\
P_R&=\frac{1}{\alpha N}\left(-L\dot{R}-\dot{L}R+(N^xRL)'\right) \label{eqPR}\\
\dot{P}_R&=-N\alpha\left(\frac{P_LP_R}{R^2}-\frac{LP_L^2}{R^3}\right)-N\frac{\md\alpha}{\md R}\left(-\frac{P_LP_R}{R}+\frac{LP_L^2}{2R^2}-\frac{L}{2}\right) \notag \\
&-\left(N\alpha\frac{R'}{L}\right)'+N'\alpha\frac{R'}{L}-\left(N'\alpha\frac{R}{L}\right)'+N'\frac{\md\alpha}{\md R}\frac{RR'}{L} \notag \\
&+\left(\frac{\md^2\alpha}{\md R^2}R+\frac{\md\alpha}{\md R}\right)\frac{NR'\,^2}{L}-\left(2N\frac{\md\alpha}{\md R}\frac{RR'}{L}\right)'+(N^xP_R)' \notag \\
&-2\pi G N P_\tau\left(1+\beta\frac{\tau'\,^2}{L^2}\right)^{-\frac{1}{2}}\frac{\tau'\,^2}{L^2}\frac{\md\beta}{\md R}\\
\dot{P}_L&=N\alpha\left(\frac{1}{2}-\frac{P_L^2}{2R^2}-\frac{R'\,^2}{2L^2}\right)-(N\alpha)'\frac{RR'}{L^2}+N^xP_L' \notag\\
&+4\pi G N\beta P_\tau\frac{\tau'\,^2}{L^3}\left(1+\beta\frac{\tau'\,^2}{L^2}\right)^{-\frac{1}{2}} \label{eqdotPL}
\end{align}
\begin{align}
\dot{\tau}&=N\left(1+\beta\frac{\tau'\,^2}{L^2}\right)^\frac{1}{2}+N^x\tau'  \label{eqTau}\\
\dot{P}_\tau&=\left[N\beta\frac{P_\tau\tau'}{L^2}\left(1+\beta\frac{\tau'\,^2}{L^2}\right)^{-\frac{1}{2}}+N^xP_\tau\right]'\,.  \label{eqPtau}
\end{align}
To try to find an LTB-like solution, following \cite{Bondi}
and the recent \cite{BKDustLTB}, we choose an embedding by coordinates
such that $\tau=t$, or equivalently, from (\ref{eqTau}),
$N=1$. Setting $N^x=0$, equation (\ref{eqPtau}) becomes
$\dot{P}_\tau=0$ so that $P_\tau(x)$ is a function of the spatial
coordinate only.

Substituting this and (\ref{eqPL}) with (\ref{eqPR}) in the
diffeomorphism constraint
\[
\frac{\delta D}{\delta N^x}=R'P_R-LP_L'+4\pi G P_\tau\tau'=0
\]
gives the equation
\[
\left(\frac{R'}{\alpha L}\right)^\centerdot=0
\]
or $R'/\alpha L=\mathcal{E}$, with $\mathcal{E}(x)$ an
arbitrary function of the spatial coordinate only. To make contact
with the classical LTB solution we may define $\kappa$ by
\begin{equation}
\mathcal{E}=\sqrt{1+\kappa}
\end{equation}
and obtain the corrected LTB condition
\[
L=\frac{R'}{\alpha\mathcal{E}}\,.
\]
Note that this is exactly the form of corrections used earlier in
(\ref{LTB2}), which we now have derived from the corrected constrained
system without extra assumptions about the possible form of LTB
solutions.

With all this, the Hamiltonian constraint 
\[
\alpha\left[-\frac{P_LP_R}{R}+\frac{LP_L^2}{2R^2}-\frac{L}{2}-\frac{R'\,^2}{2L}+\left(\frac{RR'}{L}\right)'\right]+4\pi G P_\tau=0
\]
becomes
\begin{equation}
\left[\frac{R\dot{R}^2}{\alpha^2}+R(1-\alpha^2\mathcal{E}^2)\right]'=8\pi G\mathcal{E}P_\tau \,. \label{firstOrderEq1}
\end{equation}
Again, we may define $F$ by the equation $8\pi G\mathcal{E}P_\tau=F'$, so finally the integrated Hamiltonian constraint reads
\begin{equation}
R\dot{R}^2=\alpha^2F+R\alpha^2(\alpha^2\mathcal{E}^2-1)+c(t) \label{firstOrderEq}
\end{equation}
and the second order, evolution equation from (\ref{eqdotPL}) is
\begin{equation} \label{secondOrderEq}
2R\ddot{R}+\dot{R}^2=2(\dot{R}^2+\alpha^4\mathcal{E}^2)\frac{\md\log\alpha}{\md\log R}-\alpha^2(1-\alpha^2\mathcal{E}^2)\,. 
\end{equation}
This is precisely the time derivative of (\ref{firstOrderEq}), provided $c(t)=c={\rm const}$, and shows
consistency. In the limit $\alpha=1$ we recover the classical LTB
condition and evolution equation.  In this limit the integration constant can be absorbed in $F$ so we may set $c=0$ here.

Even though the corrected LTB condition coincides with the one derived
earlier, the first order and evolution equations are
slightly different from the corresponding ones (\ref{first order eq ver two}) and (\ref{evolution eq ver two}) derived before:
\begin{equation} \label{FirstOrderSph}
\left(\frac{R\dot{R}^2}{\alpha^2}\right)'+R'(1-\alpha^2\mathcal{E}^2)-2\mathcal{E}\mathcal{E}'\alpha h R=F'
\end{equation}
which unlike (\ref{firstOrderEq1}) is not a spatial derivative,
and the second order equation
\begin{equation}\label{SecondOrderSph}
2R\ddot{R}+\dot{R}^2=2\dot{R}^2\frac{\md\log\alpha}{\md\log R}-\alpha^2(1-\alpha^2\mathcal{E}^2)\,.
\end{equation}
While several terms in the two sets of the equations agree, the
spatial derivative of (\ref{firstOrderEq}) differs from
(\ref{FirstOrderSph}) by a term $2{\cal E}{\cal E}' \alpha
R(h-\alpha)+2{\cal E}^2R\alpha\alpha'$, and (\ref{secondOrderEq}) from
(\ref{SecondOrderSph}) by $2\alpha^4{\cal E}^2 \md\log\alpha/\md\log
R$.  It is interesting to note that the more restrictive new method of
this section provides a Hamiltonian constraint which can be spatially
integrated.

\subsubsection{Case $\bar{\alpha}=1$}

Similarly to the previous case, choosing $N=1$ and $N^x=0$ gives the
equations of motion
\begin{align}
P_L=&-R\dot{R} \label{eqPL2} \\
P_R=&-(RL)^\centerdot-(1-\alpha)L\dot{R} \label{eqPR2} \\
\dot{P}_L=&\frac{\alpha}{2}+(2-\alpha)\left(-\frac{P_L^2}{2R^2}-\frac{R'\,2}{2L^2}\right) \label{dotPL} \\
\dot{P}_R=&\left(\frac{LP_L^2}{R^3}-\frac{P_LP_R}{R^2}\right)-\left(\frac{R'}{L}\right)'+\frac{\md\alpha}{\md R}\left(\frac{LP_L^2}{2R^2}+\frac{L}{2}-\frac{R'\,^2}{2L}\right) \notag \\
&+(1-\alpha)\frac{LP_L^2}{R^3}-\left((1-\alpha)\frac{R'}{L}\right)'\,.
\end{align}
Substituting the first two equations in the diffeomorphism constraint gives
\[
\left(\frac{R'}{L}\right)^\centerdot=(1-\alpha)\frac{R'\dot{R}}{RL}\,.
\]
Inserting the ansatz (\ref{LTB1})
\[
L=\frac{R'}{\mathcal{E}f}
\]
gives the same equation (\ref{diff eq for f from triads}) for the function $f(E^x)$ as in the previous
treatment:
\begin{equation}
R\frac{\md f}{\md R}=f(1-\alpha)\,.  \label{diffSol}
\end{equation}
Substituting equations (\ref{eqPL2}) and (\ref{eqPR2}) into the Hamiltonian constraint gives
\[
(2-\alpha)R'\dot{R}^2+2R\dot{R}\dot{R}'-2RR'\dot{R}\frac{\dot{f}}{f}+\alpha R' -\alpha R' \mathcal{E}^2f^2-R(\mathcal{E}^2f^2)'=8\pi G\mathcal{E}f P_\tau
\]
or using (\ref{diffSol}) and defining $F'=8\pi G\mathcal{E}P_\tau$
\begin{equation}
\alpha R'\dot{R}^2+2R\dot{R}\dot{R}'+\alpha R'(1-\mathcal{E}^2f^2)-R(\mathcal{E}^2f^2)'=fF'
\end{equation}
which reproduces the classical equation in the limit $\alpha=f=1$.

The second order equation from (\ref{dotPL}) is
\begin{equation}
2R\ddot{R}+\alpha\dot{R}^2=-\alpha(1+\mathcal{E}^2f^2)+2\mathcal{E}^2f^2 \,.
\end{equation}
Also here, compared to (\ref{first order eq ver one}) and
(\ref{evolution eq ver one}) as obtained earlier:
\[
\alpha R'\dot{R}^2+2R\dot{R}\dot{R}'+\alpha R'(1-\mathcal{E}^2f^2)-R(\mathcal{E}^2)'fl=fF'
\]
\[
2R\ddot{R}+\alpha\dot{R}^2=-\alpha(1-\mathcal{E}^2f^2) \,,
\]
the structure of the resulting equations remains similar up to a few
extra terms.

As in sections  \ref{s:effectiveDensity} and \ref{s:effectiveKappa} we may interpret the effects of correction terms using effective densities and energy functions.

\section{Possibility of singularity resolution through bounces}

We can now use the different sets of consistent equations to analyze
properties of gravitational collapse, such as the formation of
singularities. Classically we have the first order equation
\begin{equation} \label{classical first order eq}
\dot{R}^{2}=\kappa+\frac{F}{R}\,.
\end{equation}
Compared with the marginal case where $\kappa=0$, there exists the
possibility that $\dot{R}=0$ even for positive mass functions
$F$. However, to conclude whether there is a bounce or not we also
need to look at the evolution equation
\begin{equation}
2R\ddot{R}+\dot{R}^{2}=\kappa
\end{equation}
and see whether we can have $\ddot{R}>0$ in addition to
$\dot{R}=0$. From the first equation we see that $\dot{R}=0$ implies
$\kappa+F/R=0$ and with both $F$ and $R$ positive we get $\kappa<0$;
bounces would be possible only for negative $\kappa$. On the other hand,
for $\dot{R}=0$ the second order equation implies
$2R\ddot{R}=\kappa$. Since $R>0$ and $\kappa<0$ we conclude that
$\ddot{R}<0$ and thus there is no bounce classically.

We would like to proceed in a similar manner for the quantum corrected
case also. However, there the first order equation in time contains
terms with spatial derivative as well (which cannot be integrated out
in all cases). Therefore the analysis for the possibility of a bounce
cannot necessarily be done as easily as for the classical case, a
feature clearly related to the fact that we are dealing with
inhomogeneous models. Furthermore, because of the inhomogeneous nature
of the problem a bounce also makes the analysis difficult by the fact
that after the bounce there will be the possibility of shell crossing
(unless shells with larger values of $x$ bounce at larger values of
$R$). 

We note that \eqref{first order eq ver two} and \eqref{first order eq
ver one} imply, as for the classical case, that $\dot{R}=0$ is
possible in both versions of inverse triad corrections. Whether this
corresponds to a bounce is what we want to analyze. We start by
putting $\dot{R}=0$ in \eqref{first order eq ver two} which gives
\begin{equation} \label{rdot zero ver two}
(1-\alpha^{2}-\kappa\alpha^{2})R'-\kappa'\alpha hR=F'\,.
\end{equation}
Similarly, the evolution equation \eqref{evolution eq ver two} becomes
\begin{equation} \label{evolution eq for bounce ver two}
2R\ddot{R}=-\alpha^{2}(1-\alpha^{2}-\kappa\alpha^{2})\,.
\end{equation}
Using \eqref{rdot zero ver two} on the right of the above equation we have
\begin{equation} \label{ddotR}
\ddot{R}=-\alpha^{2}\frac{F'+\kappa'\alpha hR}{2RR'}\,.
\end{equation}
For a bounce, $\ddot{R}>0$ which implies that
 $-\alpha^{2}(F'+\kappa'\alpha hR)/2RR'$ should be greater than zero.
 We need to check whether this condition can be satisfied in the
 non-marginal case where, as mentioned before, there are two
 possibilities $\kappa>0$ and $-1<\kappa<0$. In what follows we will
 assume that $R'>0$, which locally around a potential bounce is a
 valid assumption even though a collapsing shell with $x=x_{1}$, say,
 would start expanding after it experiences a bounce: when
 $\dot{R}(t,x_{1})=0$ the radius of this particular shell is not
 changing with time. The assumption then essentially implies
 that a shell with $x=x_{2}>x_{1}$, which may still be contracting,
 does not immediately catch up with the $x_{1}$ shell. With this
 assumption and because $R$ and $\alpha$ are positive, the
 condition that (\ref{ddotR}) be positive becomes
\begin{equation}\label{kappa}
 \kappa'<-F'/\alpha hR 
\end{equation}
as a condition on $\kappa'$ which needs to be satisfied for a bounce,
 implying in particular that for a bounce $\kappa'$ has to be
 negative. Whether (\ref{kappa}) can be satisfied generically is not
 clear and must be determined from a numerical analysis of the
 equations.
     
For the first version of the inverse triad correction, $\dot{R}=0$ in
\eqref{first order eq ver one} gives
\begin{equation} \label{rdot zero ver one}
\alpha(1-f^{2}-\kappa f^{2})R'-\kappa'flR=fF'\,.
\end{equation}
The evolution equation \eqref{evolution eq ver one} becomes
\begin{equation} \label{evolution eq for bounce ver one}
2R\ddot{R}=-\alpha(1-f^{2}-\kappa f^{2})
\end{equation}
and using \eqref{rdot zero ver one} on the right we get
\begin{equation}
\ddot{R}=-f\frac{F'+\kappa' lR}{2RR'}\,.
\end{equation}
This should be greater than zero for a bounce. Again because of the
presence of spatial derivatives in the above expressions it is
difficult to say whether there is a bounce in general and whether or
not the singularity can be avoided.

With (\ref{firstOrderEq}) we have a corrected equation which can be
spatially integrated, allowing an analysis similar to the classical
one. The condition $\dot{R}=0$ at a bounce implies
\[
 R=\frac{F}{1-\alpha^2{\cal E}^2}= \frac{F}{1-\alpha^2-\kappa\alpha^2}
\]
which is positive provided ${\cal E}<1/\alpha$.
The second derivative
\[
 2R\ddot{R}|_{\dot{R}=0}= 2\alpha^4 \frac{\md\log\alpha}{\md\log R}{\cal E}^2- \alpha^2(1-\alpha^2{\cal E}^2)
\]
can be positive under this condition only if the derivative
$\md\log\alpha/\md\log R$ is sufficiently positive. This is not the
case in semiclassical regimes (for geometries to the right of the peak
in Fig.~\ref{LTBalpha}), where bounces are thus prohibited.  The
correction function is increasing to the left of the peak, which is a
regime with strong quantum geometry corrections. Since we have not
included all quantum corrections, a conclusion of a bounce in this
regime would be unreliable. The only semiclassical option for a bounce
is to have a geometry above the peak of inverse triad corrections, but
have decreasing patch sizes $\Delta$ which appear as the argument of
$\alpha$. Thus, the number ${\cal N}$ of patches would have to
increase sufficiently rapidly. In this regime, we have $\alpha>1$ and
thus ${\cal E}<1$ by our condition for $R>0$. Such a bounce would thus
be possible only for $\kappa<0$. (As the argument shows, without
lattice refinement a bounce from inverse triad corrections would at
best be possible only in the strong quantum regime.)

While bounces seem possible in the present situation, they cannot be
considered generic. They require a regime where the patch number is
increasing sufficiently rapidly in such a way that the patch size
decreases. Since in the discrete quantum geometry of loop quantum
gravity the patch size has a positive lower bound, the patch number of
an orbit of fixed size cannot increase arbitrarily. Tuning then seems
required to have the right behavior just when a shell is about to
bounce.

\subsection{Near center analysis regarding bounces}

On the basis of a general analysis it seems difficult to conclude
whether quantum corrections generically resolve the singularity in
non-marginal LTB models through bounces.  The simplest possibilty
seems to be one where the central shell is prevented from becoming
singular because of a bounce. For almost complete collapse, we should
expect the relevant regime to be one of small $R$. In this case the
subsequent study of the outer shells will become difficult due to
possible shell crossings, but presumably these outer shells will not
become singular either. We therefore now proceed to a near center
analysis.
   
As in the marginal case of \cite{LTB}, we use techniques similar to
those in \cite{LTBSing} and assume that near the center of the dust
cloud we can expand $R(t,x)$ as
\begin{equation} \label{series expansion for R}
R(t,x)=R_{1}(t)x+R_{2}(t)x^{2}+\cdots
\end{equation}
For the classical collapse the mass function can be expanded as
\begin{equation} \label{series expansion for F classical}
F(x)=F_{3}x^{3}+F_{4}x^{4}+\cdots
\end{equation}
Substituting the expansion for $R(t,x)$ and for $F(x)$ in the
classical first order equation $\dot{R}^{2}R=\kappa R+F$, we find that
the lowest order term on the left (as also the second term on the
right) of the above equation goes as $x^{3}$. This suggests that the
series expansion for the energy function $\kappa(x)$ should be
\begin{equation} \label{series expansion for kappa classical}
\kappa(x)=\kappa_{2}x^{2}+\kappa_{3}x^{3}+\cdots
\end{equation}
It also implies that at $x=0$, the center of the cloud, $\kappa(x)=0$
and therefore for the case where $\kappa(x)>0$ outside $x=0$,
$\kappa_{2}$ should be greater than zero. On the other hand, for
$-1<\kappa(x)<0$, $\kappa_{2}$ should be less than zero.
However, if we consider our effective $\kappa$ as in \eqref{eff kappa ver
one} then we can have lower order terms in $\kappa$. Since the lowest
order term in $\alpha$ is of order $x^{3}$ and that in $f$ is of order
$x$ we can have the lowest order term in $\kappa$ behave as
$x^{-3}$. This would then imply that at the center $\kappa$ blows up
whereas classically for negative $\kappa$ we have the condition
$-1<\kappa<0$. 

With this caveat we now consider \eqref{rdot zero ver
one} to lowest order after using various series expansions:
\begin{equation}
c_{1}R_{1}^{3}x^{3}\left(1-c_{2}^{2}R_{1}^{2}x^{2}-\frac{c_{2}^{2}\kappa_{-3}R_{1}^{2}}{x}\right)R_{1}+3c_{2}c_{3}\kappa_{-3}R_{1}^{6}x^{2}=2c_{2}F_{2}R_{1}x^{2}\,.
\end{equation}
Here $c_{1}$, $c_{2}$, $c_{3}$ and $\kappa_{-3}$ are the coefficients
of the lowest order terms in the expansion of $\alpha$, $f$, $l$ and
$\kappa$, respectively. Here we assume the orbital vertex number
${\cal N}$ to be nearly constant around the center. (This gives rise
to the strongest effect from inverse triad corrections and allows
direct comparisons with the matching results from \cite{Collapse}.)
The lowest order term in $F$ is $F_{2}$ instead of $F_{3}$ because of
the effective density correction. For $x\approx0$ the first two terms
in the parenthesis on the left can be ignored compared to the third
term and thus the above equation implies that $\dot{R}_{1}=0$ for
\begin{equation}
\kappa_{-3}=\frac{2c_{2}F_{2}}{(3c_{2}c_{3}-c_{1}c_{2}^{2})R_{1}^{5}}\,.
\end{equation}
It turns out that $3c_{2}c_{3}-c_{1}c_{2}^{2}<0$ implying that
$\dot{R}_{1}$ can be zero only for $\kappa_{-3}<0$ as in the classical
case. If we now look at \eqref{evolution eq for bounce ver one} near
the center we find that the condition for a bounce is
\begin{equation}
\kappa_{-3}>\frac{2F_{2}}{3c_{3}R_{1}^{5}}
\end{equation}
which means that, as in the classical case, $\kappa$ should be
positive implying that we do not have a bounce for the central shell.

For version two the expression for the effective density \eqref{eff
density second version} and the effective energy function \eqref{eff
kappa ver two}, respectively, imply that to lowest order the mass
function can behave as $x^{-3}$ and the energy function as
$x^{-10}$. To look at the possibility of a bounce at the center of the
cloud consider \eqref{rdot zero ver two} to lowest order:
\begin{equation}
R_{1}-c_{1}^{2}R_{1}^{7}x^{6}-\frac{\kappa_{-10}c_{1}^{2}R_{1}^{7}}{x^{4}}+\frac{10\kappa_{-10}c_{1}c_{4}R_{1}^{7}}{x^{4}}=-\frac{3F_{-3}}{x^{4}} \,.
\end{equation}
Here $c_{4}$, $F_{-3}$ and $\kappa_{-10}$ are the coefficients in the
expansion of $h$, $F$ and $\kappa$ respectively. Ignoring the first
two terms for $x\approx0$, the above equation implies
\begin{equation}
\kappa_{-10}=-\frac{3F_{-3}}{(10c_{1}c_{4}-c_{1}^{2})R_{1}^{7}}
\end{equation}
as the condition for $\dot{R}_{1}=0$. We note that the denominator
here is positive implying that $\kappa_{-10}>0$ if $F_{-3}<0$
(negative mass function near the center) and $\kappa_{-10}<0$ if
$F_{-3}>0$ (positive but decreasing mass function near the
center). None of these behaviors could occur classically; either
negative total energy ($F<0$) or a negative density ($F'<0$) would be
required. To see if the above condition implies a bounce we use
\eqref{evolution eq for bounce ver two} to lowest order and find
\begin{equation}
\kappa_{-10}>-\frac{3F_{-3}}{10c_{1}c_{4}R_{1}^{7}}
\end{equation}
as the condition for getting a bounce. This means that for $F_{-3}<0$,
$\kappa_{-10}$ has to be positive (in agreement with the condition for
$\dot{R}_{1}=0$ found above) implying that a bounce for the central
shell is possible if the above inequality is satisfied. For
$F_{-3}>0$, $\kappa_{-10}$ has to be greater than a negative number
and thus if it is positive then a bounce again seems possible.

\section{Collapse behavior near the center}
\label{s:NearCenter}

Using \eqref{series expansion for R}, \eqref{series expansion for F
classical} and \eqref{series expansion for kappa classical} we see
that to lowest order in $x$ (which is $x^{3}$) the classical equation
$\dot{R}^{2}R=\kappa R+F$ implies
\begin{equation}
\md t= \pm\frac{\md R_{1}}{\sqrt{\kappa_{2}+\frac{F_{3}}{R_{1}}}}\,.
\end{equation}
This has the solution (choosing the minus sign which corresponds to
collapse)
\begin{equation}
t=-\frac{R_{1}\sqrt{\kappa_{2}+\frac{F_{3}}{R_{1}}}}{\kappa_{2}}+\frac{F_{3}}{2\kappa_{2}^{3/2}}\log(F_{3}+2\kappa_{2}R_{1}+2\sqrt{\kappa_{2}}\sqrt{\kappa_{2}+\frac{F_{3}}{R_{1}}}R_{1})
\end{equation}
for $\kappa_{2}>0$ and 
\begin{equation}
t=\frac{R_{1}\sqrt{\frac{F_{3}}{R_{1}}-|\kappa_{2}|}}{\kappa_{2}}+\frac{F_{3}}{2|\kappa_{2}|^{3/2}}\arctan\left[\frac{\sqrt{\frac{F_{3}}{R_{1}}-|\kappa_{2}|}(2|\kappa_{2}|R_{1}-F_{3})}{2\sqrt{|\kappa_{2}|}(\kappa_{2}R_{1}-F_{3})}\right]
\end{equation}
for $\kappa_{2}<0$.

We now proceed with a similar analysis for the first version of the
inverse triad correction. Near the center the various quantities
$(\alpha, f, l)$ behave as
\begin{equation}
\alpha=\left(\frac{2}{\gamma l_{\rm P}^{2}}\right)^{3/2}R_{1}^{3}x^{3} \quad ,\quad f=\sqrt{\frac{8e^{1-\pi/2}}{\gamma l_{\rm P}^{2}}}R_{1}x \quad ,\quad l=\frac{1}{5}\left(\frac{2}{\gamma l_{\rm P}^{2}}\right)^{3/2}\sqrt{\frac{8e^{1-\pi/2}}{\gamma l_{\rm P}^{2}}}R_{1}^{4}x^{4}
\end{equation}
where the way the near center behavior for $l$ has been determined is
described around Eq.~\eqref{l near center}. In what follows we will denote the
constants appearing in the expansion of $(\alpha,f,l)$ by
$(c_{1},c_{2},c_{3})$ respectively. Thus substituting the series
expansions in \eqref{first order eq ver one} we get
\begin{equation} \label{ver one near center}
c_{1}(1-c_{2}^{2}R_{1}^{2}x^{2}-\kappa_{2}c_{2}^{2}R_{1}^{2}x^{4})R_{1}^{4}x^{3}+c_{1}\dot{R}_{1}^{2}R_{1}^{4}x^{5}+2\dot{R}_{1}^{2}R_{1}x^{2}-2\kappa_{2}c_{2}c_{3}R_{1}^{6}x^{7}=3c_{2}F_{3}R_{1}x^{3}
\end{equation}  
We now consider three different possibilities. 
\subsection{Case I: No correction to the expansion of $F$ and $\kappa$}
If we work with \eqref{ver one near center} directly then we find that the lowest order term on the left hand side is $2\dot{R}_{1}^{2}R_{1}x^{2}$ and the lowest order term on the right hand side goes as $x^{3}$ implying $\dot{R}_{1}=0$.

\subsection{Case II: Modification to the expansion of $F$ and no modification to $\kappa$}

However because of the presence of an extra factor of $x$ on the RHS we can start the expansion of $F$ with the leading term behaving as $x^{2}$. In this case the lowest order term on both the LHS and RHS are of order $x^{2}$ and we get
\begin{equation}
2\dot{R}_{1}^{2}R_{1}x^{2}=2c_{2}F_{2}R_{1}x^{2}
\end{equation}
which has the solution (for collapsing dust cloud)
\begin{equation}
R_{1}(t)=1-\sqrt{c_{2}F_{2}}(t-t_{0})
\end{equation}
where we choose the initial condition $R_{1}(t=t_{0})=1$. We see that the central singularity $R_{1}=0$ forms in a finite time $t_{s}=(1+\sqrt{c_{2}F_{2}}t_{0})/\sqrt{c_{2}F_{2}}$.

\subsection{Case III: Modifications to the series expansion of $F$ and $\kappa$}
There is a third option which is suggested by the
possibility of a corrected energy function as discussed earlier. If we
consider this correction then the lowest order term in the expansion
of $\kappa$ goes as $\kappa_{-3}/x^{3}$. With this included the
matching of lowest order terms in \eqref{ver one near center} gives
\begin{equation}
-\kappa_{-3}c_{1}c_{2}^{2}R_{1}^{5}+2\dot{R}_{1}^{2}+3\kappa_{-3}c_{2}c_{3}R_{1}^{5}=2c_{2}F_{2}
\end{equation}
and thus
\begin{equation}
\md t=-\frac{\sqrt{2}\md R_{1}}{\sqrt{2c_{2}F_{2}+(\kappa_{-3}c_{1}c_{2}^{2}-3\kappa_{-3}c_{2}c_{3})R_{1}^{5}}}
\end{equation}
with the solution
\begin{equation}
t=-\frac{\sqrt{2}R_{1}\sqrt{1+\frac{(\kappa_{-3}c_{1}c_{2}^{2}-3\kappa_{-3}c_{2}c_{3})R_{1}^{5}}{2c_{2}F_{2}}}\,{}_2\!F_1(\frac{1}{5},\frac{1}{2},\frac{6}{5},-\frac{(\kappa_{-3}c_{1}c_{2}^{2}-3\kappa_{-3}c_{2}c_{3})R_{1}^{5}}{2c_{2}F_{2}})}{\sqrt{2c_{2}F_{2}+(\kappa_{-3}c_{1}c_{2}^{2}-3\kappa_{-3}c_{2}c_{3})R_{1}^{5}}}+c_{0}
\end{equation}
where ${}_2\!F_1(a,b;c;x)$ is the hypergeometric function and
where $c_{0}$ is constant of integration.

\section{Homogeneous limit and Minkowski space}
In the classical case, the first order equation is
$\dot{R}^{2}R=\kappa(x)R+F(x)$ with corresponding expression for
the mass function $F'=8\pi GR^{2}R'\epsilon$. This allows isotropic
space-times as special solutions. We use the ansatz $R(t,x)=a(t)x$
with the condition that at time $t=0$, $a(0)=a_{0}$ and assume that
the density $\epsilon=\epsilon_{0}$ is a constant and that the energy
function goes as $\kappa=-kx^{2}$ where $k$ is a constant. When
used in the first order equation, this gives
\begin{equation} \label{classical friedmann sol}
\dot{a}^{2}a=-ka+\frac{8\pi G\epsilon_{0}}{3}
\end{equation}
which is the Friedmann equation. 

We would now like to see if we can get a Friedmann-like solution
within the LTB class with inverse triad corrections included. This
would indicate whether there can be an effective geometry of the
classical homogeneous form, although the notion of homogeneity itself
might change on a quantum space-time. Using the ansatz for $R$ and the
assumed form for the mass function and the energy function we find
that for the first version of the inverse triad correction
\eqref{first order eq ver one} we get
\begin{equation} \label{friedmann sol ver one}
\frac{\alpha(1-f^{2})}{f}a+k\alpha f^{2}ax^{2}+\frac{\alpha \dot{a}^{2}ax^{2}}{f}+\frac{2a\dot{a}^{2}x^{2}}{f}+2klax^{2}=8\pi G\epsilon_{0}x^{2}\,.
\end{equation}
Although the resulting expression is not as simple as in the marginal
case, we can see that Friedmann like solutions are not possible
since $x^2$ does not cancel from the first term while $a$ is allowed
to depend only on $t$.

The second version of the inverse triad correction \eqref{first order eq ver two} gives
\begin{equation} \label{friedmann sol ver two}
(1-\alpha^{2}+k\alpha^{2}x^{2})a+\frac{3\dot{a}^{2}ax^{2}}{\alpha^{2}}-\frac{2\dot{a}^{2}a^{2}}{\alpha^{3}}\frac{\md\alpha}{\md(ax)}+2k\alpha hax^{2}=8\pi G\epsilon_{0}x^{2}\,.
\end{equation}
Again the Friedmann solution is prohibited. Since the first term,
which spoils the homogeneous limit, is the same for an analysis based
on (\ref{firstOrderEq}), there is no homogeneous limit in that case,
either. One can also see from the first term that no other
$x$-dependent $\kappa$, which might implement quantum corrections to
the notion of homogeneity, can resolve the non-existence of
homogeneous effective geometries subject to our equations.

The homogeneous limit, as a special case, would also include Minkowski
space as the vacuum solution. Classically the first order equation for
$F=0$ and $\kappa=0$, implies that $R\equiv R(x)$ and we recover the
Minkowski spacetime. However from \eqref{first order eq ver one} we
see that after choosing the mass function and the energy function
equal to zero the equation becomes
\begin{equation}
\alpha(1-f^{2})R'+\alpha\dot{R}^{2}R'+2\dot{R}\dot{R}'R=0
\end{equation}
which, due to the presence of the first term, implies that $R$ will be
dependent on both $(t,x)$. Even though the equation has the correct
classical limit, it is not straight-forward to see how the time
dependence of $R$ should disappear in the Minkowski limit.

For the second version \eqref{first order eq ver two} gives
\begin{equation}
(1-\alpha^{2})R'+\frac{\dot{R}^{2}R'}{\alpha^{2}}+\frac{2R\dot{R}\dot{R}'}{\alpha^{2}}-\frac{2R\dot{R}^{2}R'}{\alpha^{3}}\frac{\md\alpha}{\md R}=0
\end{equation}
 which again implies that even though in the classical limit we do
 have a Minkowski solution, there is still time dependence in $R$ in
 corrected solutions.

Strong corrections are suggested at small values of the argument of
$\alpha$, which, given that $R$ determines that value, may seem
unacceptable because the center in Minkowski space is not physically
distinguished. However, the radius $R$ and thus the center is directly
relevant for the size of corrections only if there is no lattice
refinement in which case the only parameter which $\alpha$ depends on
is $R$. The primary argument of $\alpha$ is, however, the size
$\Delta$ of discrete patches rather than $R^2$ of whole spherical
orbits. With a non-trivial refinement scheme, $\alpha(R^2/{\cal N})$
will also depend on the number of vertices per orbit, which for a good
semiclassical state must provide a more uniform distribution of
quantum corrections not distinguishing a center: ${\cal N}$ must be
small when $R^2$ is small. If discrete patch sizes on all orbits are
nearly similar, quantum corrections are uniform and do not distinguish
a center. A detailed discussion would go beyond the scope of this
paper, but one can already see the crucial role played by lattice
refinement for the correct semiclassical limit.

\section{Matching of the loop corrected LTB interior with generalized Vaidya exterior}

To verify the corrected mass formulas from a different perspective, it
is instructive to match our corrected models of dust collapse to
radiative generalized Vaidya solutions.  The interior metric for an
inverse triad corrected non-marginal LTB model can be written in the
form
\begin{equation} \label{corrected ltb metric}
\md s^{2}=-\md t^{2}+\frac{(R')^{2}}{(1+\kappa)q^{2}}\md x^{2}+R^{2}\md\Omega^{2}
\end{equation}
where $q$ takes the appropriate form depending on whether the
correction function corresponds to first version or the second
version.  Following \cite{Collapse}, we  match this
interior solution with the generalized Vaidya metric
\begin{equation} \label{vaidya metric}
\md s^{2}=-\left(1-\frac{2M}{\chi}\right)\md v^{2}+2\md v\md\chi+
\chi^{2}\md\Omega^{2}
\end{equation}
where $M$ is allowed to be a function of $v$ as well as $\chi$.

Let the boundary of the dust cloud be at $x=x_{\rm b}$. Here, the exterior
coordinates will then be functions $v\equiv v_{\rm b}(t)$,
$\chi\equiv\chi_{\rm b}(t)$ of the interior coordinate $t$ such that
\begin{equation} \label{ext in terms of int}
 \md v=\dot{v}_{\rm b}\md t \quad , \quad \md\chi=\dot{\chi}_{\rm b}\md t\,.
\end{equation}
With this, the metric induced from the exterior on the matching
surface becomes
\begin{equation}
\md s^{2}=-\left(\left(1-\frac{2M}{\chi}\right)\dot{v}_{\rm b}^{2}-
2\dot{v}_{\rm b}\dot{\chi}_{\rm b}\right)\md t^{2}+\chi^{2}_{\rm b}\md\Omega^{2}
\end{equation}
Matching the first fundamental form at the boundary (where $\md x=0$
in the interior coordinates) we get
\begin{equation}
-\md t^{2}+R_{\rm b}^{2}\md\Omega^{2}=-\left(\left(1-
\frac{2M}{\chi_{\rm b}}\right)\dot{v}_{\rm b}^{2}-
2\dot{v}_{\rm b}\dot{\chi}_{\rm b}\right)\md t^{2}+
\chi^{2}_{\rm b}\md\Omega^{2}
\end{equation} 
where $R_{\rm b}$ is the value of the area radius at the boundary.
Equating the coefficients of $\md t^{2}$ and $\md\Omega^{2}$,
respectively, on the two sides implies
\begin{eqnarray} \label{first fundamental form}
\chi_{\rm b}&=&R_{\rm b} 
\nonumber \\
\left(1-\frac{2M}{\chi_{\rm b}}\right)\dot{v}_{\rm b}^{2}-
2\dot{v}_{\rm b}\dot{\chi}_{\rm b}&=&1\,.
\end{eqnarray} 

We now match the second fundamental forms (extrinsic curvature) at the
boundary. For this we need the normal to the boundary in the interior
as well as the exterior coordinates. In the interior we have $\md x=0$
which implies that the normal $n_{\mu}^{i}=(0,a,0,0)$ where $a$ is a
constant fixed by the normalization $g^{\mu\nu}_{\rm i}n_{\mu}^{\rm
  i}n_{\nu}^{\rm i}=1$ (the label $\rm i$ standing for `interior').
This gives $a=R'_{\rm b}/q_{\rm b}\sqrt{1+\kappa}$ and the normal in
the interior is $n_{\mu}^{\rm i}=(0,R'_{\rm b}/q_{\rm
  b}\sqrt{1+\kappa},0,0)$.  Evaluating the $\theta\theta$ component of
extrinsic curvature $K^{\rm i}_{\mu\nu}=n^{\rm i}_{\mu;\nu}$ we find
\begin{equation} \label{extrinsic curvature in int}
K^{\rm i}_{\theta\theta}=q_{\rm b}R_{\rm b}\sqrt{1+\kappa}
\end{equation} 
From \eqref{ext in terms of int}, at the boundary, we have $\md
v-\dot{v}_{\rm b}\md t=0$, $\md t=\frac{\md\chi}{\dot{\chi}_{\rm b}}$
and thus
\begin{equation}
 -\dot{\chi}_{\rm b}\md v+\dot{v}_{\rm b}\md\chi=0
\end{equation}
In terms of the exterior coordinates the normal to the boundary is
given by $n^{\rm e}_{\mu}=(-c\dot{\chi}_{\rm b},c\dot{v}_{\rm b},0,0)$
where $c$ is a constant fixed again by the requirement
$g^{\mu\nu}_{\rm e}n_{\mu}^{\rm e}n_{\nu}^{\rm e}=1$. Using the
inverse of the metric in the exterior we find that $c=1$. The
$\theta\theta$ component of the extrinsic curvature in the exterior is
\begin{equation} \label{extrinsic curvature in ext}
K^{\rm e}_{\theta\theta}=-\chi_{\rm b}\dot{\chi}_{\rm b}+
\chi_{\rm b}\left(1-\frac{2M}{\chi_{\rm b}}\right)\dot{v}_{\rm b}\,.
\end{equation}
Equating \eqref{extrinsic curvature in int} and \eqref{extrinsic
  curvature in ext} and using \eqref{first fundamental form} to
simplify the resulting expression we obtain
\begin{equation} \label{second fundamental form}
q_{\rm b}\sqrt{1+\kappa}=-\dot{R}_{\rm b}+\left(1-\frac{2M}{R_{\rm b}}\right)
\dot{v}_{\rm b}
\end{equation}
implying
\begin{equation}
\dot{v}_{\rm b}=\frac{q_{\rm b}\sqrt{1+\kappa}+\dot{R}_{\rm b}}{1-\frac{2M}{R_{\rm b}}}\,.
\end{equation}
Using this and its square, in the second line of \eqref{first
  fundamental form} we have
\begin{equation} \label{mass}
2M=(1-q_{\rm b}^{2}(1+\kappa)+\dot{R}_{\rm b}^{2})R_{\rm b}\,.
\end{equation}
At this stage, we can note that the expression for the Vadiya mass
resulting from matching is the same as the one obtained for the effective
Misner-Sharp mass in \eqref{ms mass first version} and \eqref{ms mass
  second version}.

The exterior region can contain trapped surfaces when the condition
$2M=\chi$ is satisfied. Using the first expression in \eqref{first
  fundamental form} along with \eqref{mass} this implies that the
boundary will be trapped when $(1-q_{\rm b}^{2}(1+\kappa)+\dot{R}_{\rm
  b}^{2})R_{\rm b} = R_{\rm b}$, or
\begin{equation} \label{condition for trapping} 
\dot{R}_{\rm b} = -q_{\rm b}\sqrt{1+\kappa}
\end{equation}
where the negative sign has been chosen since we have a collapsing
scenario. This formula shows that the horizon condition is corrected
by $q_{\rm b}$ simply by multiplying $1+\kappa^2$ as it is suggested
by the effective metric (\ref{corrected ltb metric}).

\section{Conclusions}

We have extended the treatment of \cite{LTB} to non-marginal models,
where additional corrections from spin connection terms arise. With
these additional terms the original derivation appears more arbitrary,
which led us to provide an independent derivation of equations
corrected by the treatment of inverse triads in loop quantum
gravity. In this new derivation, anomaly-freedom is implemented first
and LTB conditions are imposed afterwards to select a special class of
solutions. The structure of the resulting equations is very similar in
both derivations, showing the robustness. By the alternative method,
which is much less arbitrary than the one extended from \cite{LTB} to
non-marginal models, we thus show that the more phenomenological
treatment of corrections used in \cite{LTB} is reliable. In details,
however, the resulting equations do differ which is always possible
due to quantization ambiguities. The effects analyzed in this paper do
not appear to depend sensitively on the method, but further analysis
may well provide restrictions on acceptable equations, and thus on
quantization ambiguities.

Our analysis in this paper has been done for inverse triad
corrections, while holonomy corrections, which to some degree were
treated in the marginal case of \cite{LTB}, are technically more
involved. Already for inverse triad corrections, the extension
provided here is an interesting step in the analysis of inhomogeneous
collapse and singularities. Comparing with homogeneous models and
matching results of \cite{Collapse} would suggest easy
resolutions of singularities by bounces. Marginal models were not
entirely conclusive in this regard since their homogeneous analog is
that of a spatially flat Friedmann--Robertson--Walker model which
under inverse triad corrections gives rise to bounces only with a
negative matter potential \cite{Cyclic}. In the collapse analysis,
however, we have used positivity conditions for the mass function
which indicate that bounces in marginal LTB models with inverse triad
corrections should not be expected. For non-marginal models, on the
other hand, homogeneous special solutions with positive spatial
curvature exist, which do show bounces with inverse triad corrections
and positive matter terms \cite{BounceClosed}. One would thus expect
non-marginal models to result in bounces much more easily than
marginal ones do.

This, however, is not the case: we mostly confirm the results found in
marginal models where (i) bounces are not obvious and (ii) a
homogeneous limit of quantum corrected solutions may not even exist.
As for the first property, bounces seem somewhat easier to achieve
than in marginal models, but in contrast to the expectation turn out
to be hard to realize generically.  Moreover, a complete analysis
would have to involve an investigation of shell-crossing singularities
which can be involved even classically. (See \cite{LTB} for more
discussions on this in marginal corrected models.) As for property
(ii) about the homogeneous limit, one can evade ruling out a
homogeneous limit at the dynamical level only if one assumes a quantum
notion of symmetry which would imply effective isotropic space-time
metrics different from classical Friedmann--Robertson--Walker
models. This may well be expected, as indeed the deformed constraint
algebra (\ref{DeformedAlg}) shows that there is a corrected quantum
space-time structure. It would be interesting to see how this
influences space-time symmetries.

Finally, several issues discussed here involved the role of lattice
refinement for the semiclassical limit. As treatable models between
homogeneous ones and the full theory, LTB models turn out to be quite
instructive. This should also be expected for an implementation at the
state (rather than phenomenological) level which was started
in \cite{LTB} for marginal models but which we have not attempted here
for non-marginal models.

\section*{Acknowledgements}

We thank Tomohiro Harada for discussions. MB and JDR were supported in
part by NSF grant PHY0748336.

\end{document}